\documentclass[12pt]{article}
\usepackage{cite}
\newcommand{\be}{\begin{equation}}
\newcommand{\ee}{\end{equation}}
\newcommand{\bea}{\begin{eqnarray}}
\newcommand{\eea}{\end{eqnarray}}
\def\eqn#1{eq.~(\ref{#1})} 
\def\eqns#1#2{eqs.~(\ref{#1}) and~(\ref{#2})}

\def\ff{ f^{a_1 a_4 b} f^{a_2 a_3 b} }
\def\dd{ d^{a_1 a_4 b} d^{a_2 a_3 b} }
\def\upon{/}
\def\Tr{{\rm Tr}}
\def\e{ {\rm e} }

\def\ie{{\it i.e.}}
\def\viz{{\it viz.}}

\def\half{ {1\over 2} }

\def\A{a}
\def\ia{\alpha}
\def\ib{\beta}
\def\ic{\gamma}
\def\id{\delta}
\def\Y{-s/t}

\def\eps{\epsilon} 
\def\ep{\epsilon}
\def\reps{ r(\eps)}
\def\relleps{ r(\ell\eps)}
\def\gzell{  g_\ell}
\def\goell{  f_\ell}
\def\gz{  g_1}
\def\gone{ f_1}
\def\gztwo{  g_2}
\def\gotwo{  f_2}
\def\mommu{ { s_{ij} \over \mu^2 } }
\def\mom{ { s_{ij} \over Q^2 } }
\def\Qmu{ {Q^2 \over \mu^2 } }
\def\muQone{ \left( \mu^2 \over Q^2 \right)^{\eps} }
\def\muQtwo{ \left( \mu^2 \over Q^2 \right)^{2\eps} }
\def\muQell{ \left( \mu^2 \over Q^2 \right)^{\ell \eps} }

\def\TREE{  \left( -4 i K \over s t  \right) }
\def\regge{\quad \mathrel{\mathop{\longrightarrow}\limits_{s \gg -t}}\quad}
\def\ket#1{|{#1}\rangle}
\def\ii#1{ {[{#1}]} }
\def\ones{ \pmatrix{1\cr 1\cr 1} }
\def\tA{\tilde{A}}
\def\tAf{ \tilde{A}^{(f)} }
\def\tM{ \tilde{M} }
\def\cA{  {\cal A}  }
\def\cC{  {\cal C}  }
\def\cG{  {\cal G}  }
\def\cN{  {\cal N}  }
\def\cO{  {\cal O}  }
\def\tS{{\cal S}}
\def\tT{{\cal T}}
\def\tU{{\cal U}}
\def\bI{\mathbf{I}}
\def\bT{\mathbf{T}}
\def\bF{\mathbf{F}}
\def\bS{\mathbf{S}}
\def\bG{\mathbf{G}}

\def\bGam{\mathbf{\Gamma}}
\def\bone{1\kern -3pt \mathrm{l}}
\def\suml{\sum_{\ell=1}^\infty}
\def\sumL{\sum_{L=0}^\infty}
\def\pel{{(\ell)}}
\def\pelf{{(\ell f)}}
\def\Lp{{\ell_0}} 
\def\Ell{{(L)}}
\def\Ellf{{(Lf)}}
\def\EllL{{(L,L)}}
\def\Ellk{{(L,k)}}
\def\Ellodd{{(L,2m+1)}}

\def\Ellevenplustwo{{(L,2m+2)}}
\def\Ellzero{{(L,0)}}
\def\Ellone{{(L,1)}}
\def\Elltwo{{(L,2)}}
\def\Lpzero{{(\Lp,0)}}
\def\Zero{{(0)}}
\def\Azero{ A^\Zero} 
\def\One{{(1)}}
\def\Onezero{{(1,0)}}
\def\Oneone{{(1,1)}}
\def\Two{{(2)}}
\def\Twoone{{(2,1)}}
\def\Twotwo{{(2,2)}}
\def\TwoP{{(2)P}}
\def\TwoNP{{(2)NP}}
\def\Three{{(3)}}
\def\theequation{\thesection.\arabic{equation}}

\begin{document}
\bibliographystyle{utphys}

\begin{flushright}
BRX-TH-609\\
BOW-PH-145
\end{flushright}
\vspace{25mm}

\begin{center}
{\Large\bf\sf  
IR divergences and Regge limits of 
subleading-color contributions to 
the four-gluon amplitude in $\cN=4$ SYM Theory 
} 

\vskip 5mm Stephen G. Naculich\footnote{Research supported in part by the 
NSF under grant PHY-0756518}$^{,a}$
and Howard J. Schnitzer\footnote{Research supported in part 
by the DOE under grant DE--FG02--92ER40706\\
{\tt \phantom{aaa} 
naculich@bowdoin.edu, schnitzr@brandeis.edu}
}$^{,b}$
\end{center}

\begin{center}
$^{a}${\em Department of Physics\\
Bowdoin College, Brunswick, ME 04011, USA}

\vspace{5mm}

$^{b}${\em Theoretical Physics Group\\
Martin Fisher School of Physics\\
Brandeis University, Waltham, MA 02454, USA}
\end{center}
\vskip 2mm

\begin{abstract}

We derive a compact all-loop-order expression for the IR-divergent part
of the $\cN=4$ SYM four-gluon amplitude, which includes both planar
and all subleading-color contributions, based on the assumption that
the higher-loop soft anomalous dimension matrices are proportional to the
one-loop soft anomalous dimension matrix, as has been recently conjectured.

We also consider the Regge limit of the four-gluon amplitude,
and we present evidence that the leading logarithmic growth of the
subleading-color amplitudes is less severe than that of the planar
amplitudes.  
We examine possible $1/N^2$ corrections to the gluon Regge trajectory,
previously obtained in the planar limit from the BDS ansatz.  
The double-trace amplitudes have Regge behavior as well,
with a nonsense-choosing Regge trajectory and a Regge cut which first
emerges at three loops.

\end{abstract}

\vfil\break

\section{Introduction}
\setcounter{equation}{0}
\label{secintro}

Over the past decade, there has been much interest in 
$\cN=4$ supersymmetric SU($N$) Yang-Mills (SYM) theory, 
in part because of its relation
to string theory via the AdS/CFT correspondence, and because of the
possibility that, in the large $N$ (planar) limit, the theory may be
integrable and solvable.

Recent progress on the perturbative structure of the theory 
has been motivated by the discovery of an iterative structure 
of the loop amplitudes \cite{Anastasiou:2003kj}
which together with an analysis of IR divergences
\cite{Magnea:1990zb,Catani:1996jh,%Catani:1996vz,
Catani:1998bh, Sterman:2002qn}
led to the fruitful BDS conjecture \cite{Bern:2005iz} 
for the all-loop-orders MHV planar $n$-gluon amplitude.
This conjecture has been shown to be a consequence of dual
conformal invariance\footnote{More precisely, anomalous
dual conformal symmetry uniquely fixes the form of 
light-like Wilson loops for $n=4$ and 
$n=5$ \cite{Drummond:2006rz,Drummond:2007aua,Drummond:2007cf}, 
and much evidence has accumulated for the 
equivalence of Wilson loops to MHV planar amplitudes
\cite{Alday:2007hr,Drummond:2007aua,Brandhuber:2007yx,Drummond:2007cf,Drummond:2007bm,Bern:2008ap,Drummond:2008aq}.}
for $n=4$ and $5$, but for $n \ge 6$ must be 
modified \cite{Alday:2007he,Drummond:2007bm,Bern:2008ap,Drummond:2008aq,Anastasiou:2009kna},
though the exact form of the correction is not yet known.
In refs.~\cite{Drummond:2007aua, Naculich:2007ub,DelDuca:2008pj}
the BDS ansatz for the planar four-gluon amplitude
was shown to imply exact Regge behavior,
and the gluon Regge trajectory (in the planar limit)  was computed.
The Regge behavior of higher-point planar amplitudes has been explored in 
refs.~\cite{Brower:2008nm,Bartels:2008ce,%Bartels:2008sc,
Brower:2008ia,DelDuca:2008jg}.

While the leading-color (planar) amplitudes 
have been under intense investigation,
sub-leading-color amplitudes have received much less scrutiny.
Two-loop subleading-color four-gluon amplitudes \cite{Bern:1997nh}
can be written explicitly \cite{Smirnov:1999gc,Tausk:1999vh}
through $\cO(\eps^0)$
in a Laurent expansion in the dimensional regulator
$\epsilon = (4-D)/2$,
and three-loop subleading-color four-gluon amplitudes
are known in terms of a basis of scalar 
integrals \cite{Bern:2008pv},
but no BDS-type ansatz is known for 
general $L$-loop subleading-color amplitudes.
In previous work \cite{Naculich:2008ys}, 
we derived explicit expressions for 
the IR-divergent part of subleading-color four-gluon amplitudes
through three loops, 
and made several conjectures about the extension 
of these expressions to arbitrary loop order.
In the first part of this paper, 
we derive (using an assumption explicitly stated below)
an all-loop-orders expression for the IR-divergent part
of the four-gluon amplitude,
confirming and extending the conjectures made in ref.~\cite{Naculich:2008ys}.

The BDS ansatz was guided by an analysis of the 
IR divergences of loop amplitudes 
\cite{Magnea:1990zb,Catani:1996jh,%Catani:1996vz,
Catani:1998bh, Sterman:2002qn}.
In the planar limit, the IR divergences 
depend on two functions of the coupling:
the soft (cusp) anomalous dimension 
$\gamma(a)$ 
and the collinear anomalous dimension 
$\cG_0(a)$.
The IR divergences of subleading-color amplitudes 
depend not only on $\gamma(a)$ and $\cG_0(a)$ 
but also on a soft anomalous dimension matrix 
$\bGam(a)$. 
It was shown  \cite{MertAybat:2006wq,MertAybat:2006mz} 
that the two-loop soft anomalous dimension matrix 
is proportional to the one-loop matrix
\be
\bGam^\Two = {\gamma^\Two \over \gamma^\One} \bGam^\One
\label{GamTwo}
\ee
where $\bGam(a) = \suml a^\ell \bGam^\pel$
and $\gamma(a)  = \suml a^\ell  \gamma^\pel $.
Dixon recently established the analogous proportionality for the
matter-dependent part of the three-loop soft anomalous dimension
matrix \cite{Dixon:2009gx}.
An all-orders form for $\bGam(a)$ has been 
conjectured \cite{Bern:2008pv,Becher:2009cu,Gardi:2009qi,Becher:2009qa},
which in the case of $\cN=4$ SYM theory reduces to 
\be
\bGam^\pel
= {\gamma^\pel\over \gamma^\One} \bGam^\One
\label{Gamell} 
\ee
generalizing \eqn{GamTwo}.
In this paper, we will assume that \eqn{Gamell} holds for all $\ell$,
and thus that the $\bGam^\pel$ are mutually commuting, 
to derive a compact formula for the all-loop-order IR divergences
of the $\cN=4$ SYM four-gluon amplitude
\be
\ket{A (\eps) } 
=  \exp\left[ \suml {a^\ell \over N^\ell} \bG^\pel (\ell \eps) \right] 
\ket{H (\eps)} 
\ee
where
$ \ket{H (\eps)} $
denotes the short-distance, IR-finite, contribution to the amplitude
and 
\be
\bG^\pel (\ep) =
\frac{N^\ell}{2}
\muQone
\left[-\left(
\frac{\gamma^\pel}{\ep^2}
+\frac{2\cG_0^\pel} {\ep}  \right)\bone 
+\frac{\gamma^\pel}{4 \ep}\bGam^\One\right]
\ee
where
$\cG_0(a) = \sum_{\ell=1}^\infty  a^\ell  \cG_0^\pel$,
and $\mu$ and $Q$ are the 
renormalization and factorization scales respectively.
We use this to derive expressions for specific subleading-color
amplitudes, and to confirm and extend some of the conjectures
made in ref.~\cite{Naculich:2008ys}.

In the second part of this paper,
in an effort to see whether the iterative structures that play 
such an important role in MHV planar amplitudes might also
be present in subleading-color amplitudes,
we consider the Regge limit ($s \to \infty$ with $t$ fixed) 
of the four-gluon amplitude to all orders in perturbation theory.
We present evidence that the planar $L$-loop amplitude
has $\log^L(-s/t)$ leading log behavior,
while subleading-color amplitudes only go as $\log^{L-1}(-s/t)$,
using the IR-divergent contributions as a guide.
(To fully prove this behavior 
would require knowing the IR-finite 
parts of the amplitudes as well.)

The IR-divergent parts of the subleading-color
amplitudes possess
sufficient iterative structure to enable us (partially) 
to sum them (neglecting terms of $\cO(t/s)$)
to all orders in perturbation theory.
There are no subleading-color corrections to the
gluon Regge trajectory function through two loops, 
in agreement with the maximum transcendentality
contribution \cite{Kotikov:2002ab}
of the QCD gluon Regge trajectory
\cite{Fadin:1996tb,DelDuca:2001gu},
although the Regge residue picks up a two-loop $1/N^2$ correction.
We find Regge-type behavior for 
the IR-divergent terms of the (subleading-color) double-trace amplitudes.
It remains to be seen whether these iterative structures
extend beyond the Regge limit.

In sec.~\ref{secIR}, we derive a compact all-loop-orders expression
for the IR-divergent part of the $\cN=4$ SYM four-gluon amplitude, and
in sec.~\ref{secexpansion} we use this to derive explicit expressions
for subleading-color amplitudes.  
Section \ref{seclimit} examines the
leading logarithmic behavior of leading- and subleading-color $L$-loop
amplitudes in the Regge limit.  
In sec.~\ref{secregge}, the leading
logarithms are summed to obtain Regge trajectories.  Conclusions are
presented in sec.~\ref{secconcl}, and technical details are to be found
in two appendices.

\section{$\cN=4$ SYM IR divergences to all loops}
\setcounter{equation}{0}
\label{secIR}

In this section,
we derive a compact all-loop-orders expression 
for the IR-divergent part 
of the $\cN=4$ SYM four-gluon amplitude
in terms of anomalous dimensions 
$\gamma^\pel$ and $\cG_0^\pel$,
soft anomalous dimension matrices $\bGam^\pel$, 
and the IR-finite parts of lower-loop amplitudes.
This result relies on the assumption 
that the soft anomalous dimension matrices 
are mutually commuting, 
which follows if they are all proportional to $\bGam^\One$,
as has been recently 
conjectured \cite{Bern:2008pv,Becher:2009cu,Gardi:2009qi,Becher:2009qa}. 
We then show that our expression is consistent with previous results
at one, two, and three 
loops \cite{Catani:1998bh,Sterman:2002qn,Naculich:2008ys}.

First, we decompose the four-gluon amplitude
into a basis of traces of color generators 
\be
\cA_{4-{\rm gluon}} (1,2,3,4)
=  g^2 \sum_{i=1}^{9} A_\ii{i}  \,\, \cC_\ii{i}
\label{fourgluon}
\ee
where the color-ordered amplitudes $A_\ii{i}$
depend on the momenta $k_i$ and helicities of the gluons,
and we adopt the explicit basis 
of single and double traces  \cite{Glover:2001af} 
\bea
\label{basis}
&& \hspace{-5mm}
   \cC_\ii{1} = \Tr(T^{a_1} T^{a_2} T^{a_3} T^{a_4})\,, \qquad
   \cC_\ii{4} = \Tr(T^{a_1} T^{a_3} T^{a_2} T^{a_4})\,, \qquad
   \cC_\ii{7} = \Tr(T^{a_1} T^{a_2}) \Tr(T^{a_3} T^{a_4}) \nonumber\\
&& \hspace{-5mm}
   \cC_\ii{2} = \Tr(T^{a_1} T^{a_2} T^{a_4} T^{a_3})\,, \qquad
   \cC_\ii{5} = \Tr(T^{a_1} T^{a_3} T^{a_4} T^{a_2})\,, \qquad
   \cC_\ii{8} = \Tr(T^{a_1} T^{a_3}) \Tr(T^{a_2} T^{a_4}) \nonumber\\
&& \hspace{-5mm}
   \cC_\ii{3} = \Tr(T^{a_1} T^{a_4} T^{a_2} T^{a_3})\,, \qquad
   \cC_\ii{6} = \Tr(T^{a_1} T^{a_4} T^{a_3} T^{a_2})\,, \qquad
   \cC_\ii{9} = \Tr(T^{a_1} T^{a_4}) \Tr(T^{a_2} T^{a_3}).  \nonumber\\
\eea
Here $T^a$ are SU$(N)$ generators in the fundamental representation,
normalized according to $\Tr (T^a T^b) = \delta^{ab}$.
It is convenient to organize the color-ordered amplitudes
$A_\ii{i}$ into a vector in color space
\cite{Catani:1996jh,%Catani:1996vz,
Catani:1998bh} 
\be
\ket{A}=\left(
A_\ii{1},  \,
A_\ii{2}, \,
A_\ii{3}, \,
A_\ii{4},  \,
A_\ii{5},  \,
A_\ii{6},  \,
A_\ii{7}, \, 
A_\ii{8}, \, 
A_\ii{9}  \right)^T
\ee
where $( \cdots )^T$ denotes the transposed vector.

Next, we write the color-ordered amplitudes in a loop expansion
\be
\ket{A} = \sumL a^L \ket{A^\Ell}
\ee
where the natural 't Hooft loop expansion parameter is \cite{Bern:2005iz}
\be
a \equiv {g^2 N \over 8 \pi^2} \left( 4 \pi \e^{-\gamma} \right)^\ep \,.
\ee
Here $\gamma$ is Euler's constant,
and the loop amplitudes are evaluated using dimensional regularization
in $D= 4 - 2 \ep$ dimensions.
Although $\cN=4$ SYM theory is UV finite,
the dimensionally-regularized amplitudes
contain poles in $\ep$ due to IR divergences.
We follow the approach of 
refs.~\cite{Sterman:2002qn,MertAybat:2006mz}
to organize the IR divergences as
\be
\label{factorize}
\left|   A \left(\mommu,  a, \ep\right) \right> 
= 
 J \left(\Qmu, a, \eps \right) \, 
{\bS} \left( \mom,\Qmu,  a, \eps\right) 
\left | H \left( \mom,\Qmu,  a, \eps \right) \right>
\ee
where the prefactors $J$ and ${\bS}$ characterize the 
long-distance IR-divergent behavior, and $\ket{H}$,
which is finite as $\eps \to 0$,
characterizes the short-distance behavior of the amplitude.  
Also $s_{ij} = (k_i + k_j)^2$, 
$\mu$ is a renormalization scale,
and $Q$ is an arbitrary factorization scale
which serves to separate the long- and short-distance behavior. 
Although $Q$ was set equal to $\mu$ 
in ref.~\cite{MertAybat:2006mz}
for simplicity, 
we will keep it arbitrary. 
When we consider the Regge limit of the four-gluon amplitudes
in secs.~\ref{seclimit} and \ref{secregge}, 
we will set $Q^2$ equal to the fixed momentum scale $-t$. 

Because $\cN=4$ SYM theory is conformally invariant,
the product of jet functions $J$ 
may be explicitly evaluated as \cite{Bern:2005iz}
\be
J \left( \Qmu, a, \eps \right) 
= \exp \left[ -\frac{1}{2} \sum_{\ell=1}^\infty a^\ell
\muQell
\left( { \gamma^\pel \over  (\ell \eps)^2 }
        + { 2 \cG_0^\pel \over  \ell \eps }
\right)
\right]
\ee
where 
$\gamma^\pel$ and $\cG_0^\pel$ are the coefficients of 
the soft (or Wilson line cusp) and collinear anomalous dimensions 
of the gluon respectively 
\bea
\label{defanom}
\gamma (a)
&=& 
\sum_{\ell=1}^\infty a^\ell  \gamma^\pel = 
4 a - 4 \zeta_2 a^2 + 22 \zeta_4 a^3 + \cdots
\nonumber\\
\cG_0(a)
&=& 
\sum_{\ell=1}^\infty  a^\ell  \cG_0^\pel =
    -  \zeta_3 a^2 + (4 \zeta_5 + \frac{10}{3} \zeta_2 \zeta_3 )  a^3 + \cdots
\eea
The soft function ${\bS}$,
written in boldface to indicate that it is a matrix acting on the vector
$\ket{H}$,
is given by \cite{Sterman:2002qn,MertAybat:2006mz}
\be
{\bS} \left( \mom, \Qmu, a, \eps \right) 
\,=\,
{\rm P}~{\rm exp}\left[
\, -\; \frac{1}{2}\int_{0}^{Q^2} \frac{d\tilde{\mu}^2}{\tilde{\mu}^2}
\bGam \left( \mom,
             {\bar a}  \left(\frac{\mu^2}{\tilde{\mu}^2}, a, \eps  \right)
       \right) 
\right]\,
\label{soft}
\ee
where\footnote{We suppress the explicit dependence of $\bGam^\pel$ 
on ${ s_{ij} / Q^2 }$ to lighten the notation.}
\be
\bGam \left( \mom, a \right)
= \suml a^\ell \bGam^\pel, \qquad\qquad
{\bar a}  \left(   \frac{\mu^2}{\tilde{\mu}^2}, a, \eps \right) 
=  \left(     \frac{\mu^2}{\tilde{\mu}^2}   \right)^\eps  a.
\ee
The integral (\ref{soft}) is path-ordered,
but this becomes irrelevant
if the soft anomalous dimension matrices $\bGam^\pel$
all commute with one another.
In ref.~\cite{MertAybat:2006wq,MertAybat:2006mz} 
it was shown that 
$ \bGam^\Two = {1 \over 4} \gamma^\Two \bGam^\One $ ,
and in ref.~\cite{Dixon:2009gx} that 
$ \bGam^\Three = {1 \over 4} \gamma^\Three \bGam^\One$
for the non pure gluon contributions. 
If we assume that
\be 
\bGam^\pel = {\gamma^\pel\over 4}  \bGam^\One
\qquad {\it (assumption)}
\label{prop}
\ee
holds for all $\ell$ in $\cN=4$ SYM theory,\footnote{
Difficulties may arise at four loops, however,
due to the possibility of quartic Casimir 
terms \cite{Gardi:2009qi,Dixon:2009gx,Armoni:2006ux,Alday:2007mf,Dixon}.}
as has been conjectured in 
refs.~\cite{Bern:2008pv,Becher:2009cu,Gardi:2009qi,Becher:2009qa},
then the $\bGam^\pel$ indeed commute,\footnote{The 
assumption that $\bGam^\pel$ commute
was also used to simplify the
IR divergences of QCD in ref.~\cite{Becher:2009qa}.}
and we can explicitly integrate
\eqn{soft} to obtain
\be
{\bS} \left( \mom, \Qmu, a, \eps \right) 
 = \exp \left[ \frac{1}{2} \sum_{\ell=1}^\infty a^\ell
\muQell
{ \bGam^\pel   \over  \ell \eps }
\right].
\ee
Combining the exponents of the jet and soft functions 
into\footnote{In ref.~\cite{Naculich:2008ys}, 
$Q$ was set equal to $\mu$.} \cite{Sterman:2002qn,Naculich:2008ys}
\be
\bG^\pel (\ep) =
\frac{N^\ell}{2}
\muQone
\left[-\left(
\frac{\gamma^\pel}{\ep^2}
+\frac{2\cG_0^\pel} {\ep}  \right)\bone 
+\frac{1}{\ep}\bGam^\pel\right]
\label{defG}
\ee
we may express the four-gluon amplitude 
in the compact 
form\footnote{Henceforth we suppress 
$ s_{ij} $, $Q$, $\mu$, and $a$
in the arguments of the amplitudes.}
\be
\label{compact}
\ket{A (\eps) } 
=  \exp\left[ \suml {a^\ell \over N^\ell} \bG^\pel (\ell \eps) \right] 
\ket{H (\eps)} 
\ee
which will be very useful in extracting the IR-divergent parts of
subleading-color amplitudes in sec.~\ref{secexpansion}.
The expression (\ref{compact}) is valid
up to the number of loops $L$ 
for which the set of soft anomalous dimension matrices 
$\{ \bGam^\pel ~|~ \ell \le L\}$
mutually commute, 
at least $L=2$ and possibly to all orders.

We now briefly show that \eqn{compact} is consistent with previous results 
at one, two, and three 
loops \cite{Catani:1998bh,Sterman:2002qn,Naculich:2008ys}.
Equations~(3.13-3.15) of ref.~\cite{Naculich:2008ys} 
and their generalization to all $L$ are compactly written as
\be
\label{deffinite}
\ket{ \tAf (\eps)} 
= \sumL a^L \ket{\tA^\Ellf(\eps)} 
= \left( \bone - \suml \frac{a^\ell }{N^\ell}\bF^\pel (\ep)   \right) 
\ket{A (\eps) }.
\ee
The $\bF^\pel$ are chosen so as to cancel all the IR 
divergences in $\ket{A (\eps) }$, leaving an IR-finite expression 
$\ket{\tAf(\eps)}$. 
In view of \eqn{compact}, this can be accomplished by requiring
\be
\label{defF}
\left( \bone - \suml \frac{a^\ell }{N^\ell} \bF^\pel (\ep)  \right) 
\exp\left[ \suml \frac{a^\ell }{N^\ell}\bG^\pel (\ell \eps)   \right] = \bone \,.
\ee
The $\bF^\pel$ defined by \eqn{defF} may be written more explicitly as follows.
In ref.~\cite{Bern:2005iz},  the functional $X[M]$ was defined via
\be
\label{defX}
1 + \suml a^\ell M^\pel 
\equiv \exp \left[ \suml a^\ell \left( M^\pel - X^\pel[M] \right)\right]
\ee
thus, e.g., 
$X^\One[M] = 0 $,
$X^\Two[M] = {1\over 2} \left[M^\One \right]^2 $,
$X^\Three[M] = -{1\over 3} \left[M^\One \right]^3
+ M^\One M^\Two $, etc.
This functional was defined for scalar functions $M^\pel$,   
but we can also use it for commuting matrices.
We have assumed that $\bGam^\pel$ and therefore $\bG^\pel$ 
all commute with one another, and thus 
$\bF^\pel$ do so as well as a consequence of \eqn{defF}.
Thus we can write
\be
\left( \bone - \suml \frac{a^\ell}{N^\ell}  \bF^\pel (\ep)  \right) 
= \exp \left[ \sumL \frac{a^\ell}{N^\ell}  
\left(- \bF^\pel (\eps) - X^\pel[-\bF] \right)\right]
\ee
and so \eqn{defF} is equivalent to 
\be
\bF^\pel  (\eps) = -  X^\pel[-\bF]  + \bG^\pel (\ell \eps) 
\label{recurse}
\ee
which defines $\bF^\pel$ recursively in terms of 
$\bG^\pel$ and $\bF^{(\ell')}$ with $\ell' < \ell$.
Equation (\ref{recurse}) precisely agrees,
in the case where the $\bF^\pel$ commute with one another,
with eqs.~(3.16-3.18) of ref.~\cite{Naculich:2008ys}\footnote{
Based on the results of ref.~\cite{Sterman:2002qn}.}
for $\ell \le 3$, 
and provides their all-orders generalization.
Equations (\ref{compact}--\ref{defF}) then imply
\be
\ket{\tAf(\eps)}  = \ket{H (\eps)} 
\ee
that is, the IR-finite function defined via \eqn{deffinite}
is identical to the short-distance function defined in \eqn{factorize}.

In appendix \ref{appa}, we show how \eqn{compact} may also be used
to easily obtain the 
IR-divergent part of
the $L$-loop generalization \cite{Bern:2005iz} 
of the ABDK equation \cite{Anastasiou:2003kj}.

\section{IR divergences in the $1/N$ expansion} 
\setcounter{equation}{0}
\label{secexpansion}

The $L$-loop color-ordered amplitudes 
may be written in a $1/N$ expansion as
\be
\ket{A^\Ell(\eps)} = \sum_{k=0}^L \frac{1}{N^k} \ket{A^\Ellk(\eps)}
\ee
where $\ket{A^\Ellzero}$ are the leading-color (planar) amplitudes and 
$\ket{A^\Ellk}$,  $1 \le k \le L$,
are the subleading-color amplitudes.
The $L$-loop planar amplitudes are predicted by the 
BDS ansatz \cite{Bern:2005iz},
but no general expression is known for the $L$-loop
subleading-color amplitudes
(although exact expressions in terms of 
scalar integrals are known through three loops \cite{Bern:2008pv}). 
In this section, we will use the result (\ref{compact})
derived in sec.~\ref{secIR}  to extract explicit
expressions for the IR-divergent parts of subleading-color amplitudes.
These will be useful in discussing the Regge limits of these
amplitudes in secs.~\ref{seclimit} and \ref{secregge}.

We begin by expanding \eqn{compact}:
\be
\hspace{-1mm}
\ket{A(\eps)} = 
\sumL \sum_{k=0}^L  \frac{a^L}{N^k} \ket{A^\Ellk(\eps)} = 
\prod_{\ell=1}^\infty 
\sum_{\{n_\ell\}}  
{1 \over n_\ell !}
\left( a^\ell \frac{\bG^\pel (\ell \eps) }{N^\ell}  \right)^{n_\ell} 
\sum_{\ell_0=0}^\infty \sum_{k_0 = 0}^{\ell_0} {a^{\ell_0} \over N^{k_0}} 
\ket{H^{(\ell_0, k_0)}(\eps)}.
\label{expandinit}
\ee
Assuming that the proportionality (\ref{prop}) holds, 
we use \eqn{defG} to write
\be
\label{newG}
\frac{\bG^\pel (\ell \ep)} {N^\ell}
 = 
  {1 \over 2}
\muQell
\left[-\left(
\frac{\gamma^\pel}{(\ell \ep)^2}
+\frac{2\cG_0^\pel} {\ell \ep}  \right)\bone 
+\frac{\gamma^\pel}{4 \ell \ep}\bGam^\One\right].
\ee
The one-loop soft anomalous dimension matrix,
defined by \eqn{oneloopanom}, takes the form
\be
\label{defGamOne}
\bGam^\One = 
2 \left( \begin{array}{cc}
 \ia &     0 \\
    0  &  \id 
\end{array}
\right)
+ {2 \over N} \left( \begin{array}{cc} 0 & \ib \\
\ic  & 0
\end{array}
\right)
\ee 
where explicit expressions for the momentum-dependent matrices 
$\ia$, $\ib$, $\ic$,  and $\id$
are given in appendix \ref{appb}.
Due to the assumption (\ref{prop}),
the $1/N$ expansion of ${\bG^\pel (\ell \eps) }/{N^\ell} $
has only two terms
\be
\frac{\bG^\pel (\ell \eps) }{N^\ell}   
=
\gzell  + {1 \over N} \goell
\ee
where $\gzell$ and $\goell$ can be read from \eqns{newG}{defGamOne}.
We rewrite \eqn{expandinit}  as
\be
\hspace{-1mm} 
\ket{A(\eps)} = 
\sumL \sum_{k=0}^L  \frac{a^L}{N^k} \ket{A^\Ellk(\eps)} = 
\prod_{\ell=1}^\infty 
\sum_{\{n_\ell\}}  
{1 \over n_\ell !}
\left( a^\ell \gzell + {a^\ell \over N} \goell  \right)^{n_\ell} 
\sum_{\ell_0=0}^\infty \sum_{k_0 = 0}^{\ell_0} {a^{\ell_0} \over N^{k_0}} 
\ket{H^{(\ell_0, k_0)}(\eps)}
\label{expand}
\ee
so that all $N$ dependence is explicit.

Now consider an individual term on the r.h.s.~of \eqn{expand}.
By counting powers of $a$ and $1/N$, one sees that 
this term contributes to $\ket{A^\Ellk(\eps)}$, with
\be
\label{defLk}
L= \ell_0 + \suml   \ell n_\ell  , \qquad
k=k_0+k_1
\ee
where $k_1$ is the number of factors $\goell$ present in the term.
{}From \eqns{newG}{defGamOne},
it is apparent that $\gzell$ has a double pole in $\eps$,
but $\goell$ only has a single pole. 
The leading IR pole in the term under consideration is 
therefore $1/\eps^p$, where
\be
p =2 \suml n_\ell - k_1\,.
\label{firstp}
\ee
Combining \eqns{defLk}{firstp}, 
we find
\be
p = 2 L - k 
- \left[ 2 \suml (\ell -1) n_\ell  + 2 \ell_0 - k_0 \right].
\label{defp}
\ee
Since $k_0 \le \ell_0$, 
the term in square brackets is non-negative, so the leading IR
pole of $\ket{A^\Ellk (\eps)}$ is
\be
\ket{A^\Ellk(\eps)}  \sim \cO \left( \frac{1}{\eps^{2L-k}}\right).
\ee
This behavior was previously established in ref.~\cite{Naculich:2008ys} for 
amplitudes through $L=3$.

\subsection{Leading IR divergence of $A^\Ellk$}

We now derive the coefficient of the
leading IR pole of $\ket{A^\Ellk(\eps)}$.
Terms in \eqn{expand} contribute to the leading
IR pole only when the expression in square brackets in \eqn{defp}
vanishes,
which occurs when $n_\ell=0$ for $\ell \ge 2$, and $\ell_0=k_0=0$
(with $n_1$ unconstrained).     
In other words, the leading IR divergences are given by
\be
\ket{A(\eps)}  
\sim  \exp\left[  a \frac{\bG^\One (\eps) }{N}  \right] \ket{\Azero}
\qquad {\it (leading~IR~divergence)}
\label{leading}
\ee
where $\ket{H^{(0,0)}}  = \ket{\Azero}$.  
This confirms a conjecture\footnote{
In that paper we expressed this in terms of
$\bI^\One$, the operator introduced in 
ref.~\cite{Catani:1996jh,%Catani:1996vz,
Catani:1998bh},
but as we showed there $\bI^\One$ and $\bG^\One$ only differ by terms
subleading in $\epsilon$.}
made in ref.~\cite{Naculich:2008ys}.
Recalling that
\be
\frac{\bG^\One (\eps) }{N}   
= \muQone
\left[ - {2 \over \eps^2} \bone 
    + {1 \over \eps}
\left( \begin{array}{cc}
\ia &   0   \\
  0    & \id 
\end{array} \right)
+ {1 \over N \eps}
\left( \begin{array}{cc}
  0 & \ib   \\
  \ic   & 0
\end{array} \right)
\right]
\ee
we use \eqn{leading} to obtain 
the coefficient of the leading IR pole
\be
\ket{A^\Ellk (\eps)} = 
{(-2)^{L-k} \over k! (L-k)!}  
{1 \over \eps^{2L-k}}
\left( \begin{array}{cc}
  0 & \ib   \\
  \ic   & 0
\end{array} \right)^k \ket{\Azero}
+ \cO\left( 1 \over \eps^{2L-k-1} \right).
\ee
The leading IR pole of the planar amplitude is simply
\be
\ket{A^\Ellzero (\eps)} = 
{(-2)^{L} \over L! ~\eps^{2L}}  \ket{\Azero}
+ \cO\left( 1 \over \eps^{2L-1} \right)
\ee
with the rest of the IR divergences
given by the (generalized) ABDK equation (see appendix \ref{appa}).
The leading IR poles of the subleading-color amplitudes
may be written explicitly using eqs.~(\ref{defalpha}-\ref{defXYZ}),
\be
\ket{A^\Ellodd (\eps)}
=
\left({-4 i K \over stu} \right)
{(-1)^{L-1} 2^{L-m}  
\left( X^2 + Y^2 + Z^2 \right)^m 
(sY-tX)
\over 
(2m+1)! (L-2m-1)!  
\,\eps^{2L-2m-1}}
\pmatrix{
0\cr
0\cr
0\cr
0\cr
0\cr
0\cr
1\cr
1\cr
1\cr}
+ 
\cO\left( 1 \over \eps^{2L-2m-2} \right)
\label{odd}
\ee
and
\be
\ket{A^\Ellevenplustwo (\eps)}
=
\left({-4 i K \over stu} \right)
{
(-1)^{L} 2^{L-m-1}  
\left( X^2 + Y^2 + Z^2 \right)^m 
(sY-tX)
\over 
(2m+2)! (L-2m-2)!
\eps^{2L-2m-2}
}  
\pmatrix{
X-Y\cr
Z-X\cr
Y-Z\cr
Y-Z\cr
Z-X\cr
X-Y\cr
0\cr
0\cr
0 \cr}
+
\cO\left( 1 \over \eps^{2L-2m-3} \right)
\label{even}
\ee
where $s$, $t$, and $u$ are the Mandelstam invariants,
$K$ depends on the momenta and helicity of the gluons, 
and is totally symmetric under permutations of the external legs,
and 
\be
X = \log \left(t \over u\right), \qquad
Y = \log \left(u \over s\right), \qquad
Z = \log \left(s \over t\right).
\ee
The results (\ref{odd}) and (\ref{even})
are generalizations of the expressions derived in ref.~\cite{Naculich:2008ys}.

\subsection{IR divergences of $A^\EllL$}

In the previous section, 
we derived the coefficient of the leading IR pole 
of the leading- and subleading-color amplitudes $\ket{A^\Ellk}$.
It is also possible to use \eqn{expand} 
to derive further terms in the Laurent expansion.

In this section, 
we derive an expression 
for the IR divergences of the 
most subleading-color amplitude $\ket{A^\EllL}$.
The only terms in \eqn{expand} that contribute to $\ket{A^\EllL}$
are those with as many factors
of $1/N$ as of $a$. 
Thus, only $\gone$ and $ \ket{H^{(\ell_0, \ell_0)}}$ can contribute,
giving 
\be
\hspace{-1mm}
\ket{A^\EllL(\eps) } 
= \sum_{\ell_0=0}^L   {1\over (L-\ell_0)!} 
\gone^{L-\ell_0}
\ket{H^{(\ell_0,\ell_0)}(\eps)},
\qquad
{\rm where}
\quad
\gone
=
{1 \over \eps}
\muQone
\left( \begin{array}{cc}
  0 & \ib   \\
  \ic   & 0
\end{array} \right)
\ee
exact to all orders in the $\eps$ expansion.
Keeping just the first two terms in the Laurent expansion, we find
\bea
\ket{A^\EllL(\eps)}
 &=& 
\frac{1}{(L-1)!} 
\gone^{L-1}
\left[ \frac{1}{L} \gone \ket{\Azero} 
+ \ket{H^{(1,1)}(\eps)}  \right]
+ \cO\left( 1 \over \eps^{L-2} \right)
\nonumber
\\ 
&=&
\frac{1}{(L-1)!} 
{1\over \eps^{L-1}} 
\left( \begin{array}{cc}
  0 & \ib   \\
  \ic   & 0
\end{array} \right)^{L-1}
\ket{A^{(1,1)}(L \eps)}  + \cO\left( 1 \over \eps^{L-2} \right).
\label{mostsubleading}
\eea
This confirms the conjecture made in eqs.~(4.45) and (4.46) 
of ref.~\cite{Naculich:2008ys}.

\subsection{IR divergences of $A^\Ellone$}
\label{secIREllone} 

In this section, we consider the subleading-color amplitude 
$\ket{A^\Ellone}$,
and derive the first 
three\footnote{It is straightforward to obtain further terms 
in the Laurent expansion as needed.}
terms in the Laurent expansion.
Consider all terms in \eqn{expand}
for which the expression in square brackets in \eqn{defp}
is $\le 2$:
\bea
\ket{A^\Ell(\eps)} 
&=& \frac{1}{L!} 
    \left( \gz + {1\over N} \gone \right)^L \ket{\Azero}
+ \frac{1}{N (L-1)!} 
    \left( \gz + {1\over N} \gone       \right)^{L-1} 
						\ket{H^{(1,1)}(\eps)}
\label{firstthree}
\\
&+&
 \frac{1}{ (L-2)!} 
    \left( \gz + {1\over N} \gone       \right)^{L-2} 
    \left( \gztwo + {1\over N} \gotwo          \right) \ket{\Azero}
+ \frac{1}{ (L-1)!} 
    \left( \gz + {1\over N} \gone       \right)^{L-1} 
						\ket{H^{(1,0)}(\eps)}
\nonumber\\
&+&
 \frac{1}{ N^2 (L-2)!} 
    \left( \gz + {1\over N} \gone       \right)^{L-2} 
						\ket{H^{(2,2)}(\eps)}
+ \cdots
\qquad\qquad {\it (three~leading~IR~poles)}
\nonumber
\eea
where we use \eqns{newG}{defGamOne} to write 
\bea
\gz
= \muQone \left[- {2 \over \eps^2} \bone 
  + {1 \over \eps}
\left( \begin{array}{cc}
\ia &   0   \\
  0    & \id 
\end{array} \right) \right],
&&
\gone
=
{1 \over \eps}
\muQone
\left( 
\begin{array}{cc}
  0 & \ib   \\
  \ic   & 0
\end{array} \right),
\nonumber\\
\gztwo=
\muQtwo \left[
 - \left( {\gamma^\Two \over 8 \eps^2} + {\cG_0^\Two \over 2 \eps} \right)
 \bone 
+ {\gamma^\Two \over 8 \eps} 
\left( \begin{array}{cc}
\ia &   0   \\
  0    & \id 
\end{array} \right) \right],
&&
\gotwo
=
{\gamma^\Two \over 8 \eps} 
\muQtwo
\left( 
\begin{array}{cc}
  0 & \ib   \\
  \ic   & 0
\end{array} \right).
\qquad
\qquad
\label{onetwo}
\eea
To extract the $\ket{A^\Ellone}$ amplitude,
we employ the identity
\bea
&&\left( \gz + {1\over N} \gone \right)^L  \Bigg|_{1/N~{\rm piece}}
\label{ident}
\\
&& \quad=
L \gz^{L-1} \gone
~+~ {L \choose 2} \gz^{L-2} [\gone,\gz]
~+~ {L \choose 3} \gz^{L-3} [ [\gone,\gz], \gz]
%+ {L \choose 4} \gz^{L-4} [ [ [\gone,\gz], \gz], \gz] 
~+~  [ \cdots [[[\gone,\gz], \gz], \gz]  \cdots ]
\nonumber
\eea
in which the first term on the r.h.s. has an expansion 
that starts with $1/\eps^{2L-1}$,
the second term has an expansion 
that starts with $1/\eps^{2L-2}$,
and so forth.
Thus, keeping only the 
terms proportional to $1/N$ in \eqn{firstthree},
and only the first three terms in the Laurent expansion,
we obtain
\bea
\ket{A^\Ellone} 
&=& 
\frac{1}{(L-1)!} \gz^{L-1} \gone \ket{\Azero}
+ \frac{1}{2 (L-2)!} \gz^{L-2} [\gone,\gz] \ket{\Azero}
+ \frac{1}{(L-1)!} \gz^{L-1} \ket{H^{(1,1)}(\eps)} 
\nonumber\\
&+&
 \frac{1}{6(L-3)!} \gz^{L-3} [ [\gone,\gz], \gz] \ket{\Azero}
+ \frac{1}{ (L-2)!} 
    \gz^{L-2} \gotwo \ket{\Azero}
+ \frac{1}{ (L-3)!} 
    \gz^{L-3} \gone \gztwo  \ket{\Azero}
\nonumber\\
&+&
 \frac{1}{ (L-2)!}  \gz^{L-2} \gone \ket{H^{(1,0)}(\eps)}
+ \cO \left( \frac{1}{\eps^{2L-4}} \right) .
\label{firstsubleading}
\eea

In sec.~\ref{secdoubletraj}, 
we will study the subleading-color amplitude $\ket{A^\Ellone}$
in the Regge limit $s \gg -t$, with $t<0$ held fixed. 
In anticipation of that, we now compute the Regge limit of
the IR-divergent expression (\ref{firstsubleading}), 
neglecting terms suppressed by powers of $t/s$.
It is convenient in the Regge limit
to choose the factorization scale $Q^2$ 
equal to the (fixed) momentum scale $-t$.
Thus, using \eqn{onetwo} together with \eqns{tree}{defnewalpha}
we obtain
\bea
\ket{A^\Ellone(\eps)} 
&=& 
\TREE
\left(\mu^2 \over -t\right)^{L\epsilon}  
\frac{(-2)^L}{(L-1)!}  \frac{Y}{\eps^{2L-1} }
\Bigg[  
  \pmatrix{
 1\cr 1\cr 1\cr}
+ \frac{3 (L-1)}{4} \eps
  \pmatrix{
 -Z \cr X  \cr 0\cr}
\label{tedious} \\
&+& \frac{(L-1)(L-2)}{24} \eps^2 
  \pmatrix{
 7 Z^2 \cr 7 X^2 \cr  X Z \cr}
+ \frac{(L^2-17L+12) \zeta_2}{8} \eps^2
  \pmatrix{
 1  \cr 1  \cr 1  \cr }
+ \cO (\eps^3)  
+ \cO \left( t\upon s  \right) 
\Bigg]
\nonumber
\eea
% Though the above equation was only derived for L>1
% it is also valid for L=1 because previous equation is exact for L=1
where we have suppressed the first six (vanishing) entries of the vector.
To obtain \eqn{tedious}, we also needed to use 
terms through $\cO(\eps)$ in 
\be
\ket{H^\Oneone (\eps)} 
=
\TREE
\left[ \zeta_2 Y \eps + \cO(\eps^2) 
+ \cO \left( t\upon s  \right) 
\right]
\pmatrix{ 1  \cr 1  \cr 1  \cr }
\label{finiteone}
\ee
as well as the $\eps\to 0$ limit of  $\ket{H^\Onezero(\eps)}$, namely 
\be
\ket{H^\Onezero (0)} 
= 
\TREE
\left[ 4 \zeta_2  \left( 1, 0, -1, - 1, 0,  1,  0,0,0\right)^T 
+ \cO \left( t \upon s  \right) 
\right]
\label{finitezero}
\ee
which are obtained from the Laurent expansions of the
exact expressions (\ref{defM}) and (\ref{oneone}).
(Note that terms suppressed by powers of $t/s$ have
been omitted in both \eqns{finiteone}{finitezero}.)

\section{Regge limit of  $\cN=4$ SYM four-gluon amplitudes} 
\setcounter{equation}{0}
\label{seclimit}

In this section, we consider the leading logarithmic 
behavior of $L$-loop planar and subleading-color 
$\cN=4$ SYM four-gluon amplitudes
in the Regge limit $ s \gg -t $, with $t<0$ held fixed.
In sec.~\ref{secregge}, we sum the leading logs 
to obtain the Regge trajectories.

\vfil\break

\subsection{Expectations from transcendentality}

The $L$-loop planar and subleading-color amplitudes 
may be written as
\be
\ket{A^\Ellk(\eps)}  =  
\TREE
\left(\mu^2 \over -t\right)^{L\epsilon}  
\sum_{m=-2L+k}^\infty  \eps^m  \, \ket{\A^\Ellk_m (s/t)}
\ee
where
$\ket{\A^\Ellk_m (s/t)}$ is generally a 
complicated function of logarithms and polylogarithms.
(We consider the amplitude in the physical region
$s>0$ and $t,u<0$,  with $s+t+u=0$.)
All $\cN=4$ SYM amplitudes have been observed to have uniform 
transcendentality [\citen{Kotikov:2002ab,Bern:2006ew},\citen{Naculich:2008ys}]. 
This means that $\ket{\A^\Ellk_m (s/t)}$ 
is a function of $s/t$ whose degree of 
transcendentality\footnote{Each factor of 
$\zeta_k$, $\pi^k$, $\log^k (-s/t)$,
or any polylogarithm of total degree $k$ has transcendentality $k$,
and the transcendentality of a product of factors is additive.}
is $2L+m$.

Now we consider 
$\ket{\A^\Ellk_m (s/t)}$ 
in the Regge limit $ s \gg -t $, with $t<0$ held fixed.
Dropping any terms suppressed by at least one power of $t/s$,
we are left with a polynomial in $\log(-s/t)$.
Since logarithms have unit transcendentality, 
the degree of the polynomial can be no greater than $2L+m$.
In the Regge limit, 
$\ket{\A^\Ellk_m (s/t)}$ will be dominated
by the leading term in the polynomial.
{\it A priori} we might expect 
this term to be the maximum allowed by transcendentality,  
so that
\be
\ket{\A^\Ellk_m (s/t)}
\regge
{\rm const} \left[\log\left(-{s\over t}\right)\right]^{2L+m}
+ {\rm subleading}
\qquad ({\it a~priori~expectation})
\label{expectation}
\ee
where ``subleading'' indicates that we have dropped 
lower powers of $\log(-s/t)$ as well
as terms suppressed by powers of $t/s$.

The expectation (\ref{expectation}), however, is incorrect;
the leading power of $\log(-s/t)$ is almost always 
less than the maximum allowed by transcendentality.
The evidence suggests that the Regge limit
of the planar $L$-loop amplitude 
is given by\footnote{Terms suppressed by powers of $t/s$,
however,  can, and do, 
contain powers of $\log(-s/t)$ higher than $L$.}
\be
\ket{\A_m^\Ellzero (s/t)}
\regge
c_{L+m} \left[\log \left(-{s \over t}\right)\right]^{L} 
+ {\rm subleading}
\qquad {\it(conjecture)}
\label{planarconj}
\ee
where $c_{L+m}$ is a constant with degree of transcendentality $L+m$
(and vanishes for $m<-L$, 
in which case the lower powers of $\log(-s/t)$ cannot be neglected).
The leading logarithmic growth 
of subleading-color amplitudes in the Regge limit 
appears to be even weaker than that for planar amplitudes,
and we conjecture that
\be
\ket{\A_m^\Ellk (s/t)}
\regge
c'_{L+m+1}
\left[\log \left(-{s \over t}\right)\right]^{L-1} 
+ {\rm subleading},
\qquad 
{\rm for~}k\ge 1 
\qquad {\it(conjecture)}
\label{nonplanarconj}
\ee
where $c'_{L+m+1}$ is a constant with degree of transcendentality $L+m+1$
(and vanishes when $m<-L-1$).
We will discuss the evidence for the claims (\ref{planarconj})
and (\ref{nonplanarconj}) in the remainder of this section.

\subsection{Regge limit of planar amplitudes}
\label{secplanarlimit}

In this section, we review the Regge limit of the BDS ansatz for 
the planar four-gluon amplitude,
which was explored in 
refs.~\cite{Drummond:2007aua, Naculich:2007ub,DelDuca:2008pj}.

The BDS ansatz for $A^\Ellzero_\ii{1}$ is \cite{Bern:2005iz}
\bea
A^\Ellzero_\ii{1} 
&=&
M^\Ell (s,t; \eps)  ~ A^\Zero_\ii{1},
\\
A^\Zero_\ii{1} 
&=&
- {4 i K \over s t  },
\\
1 + \sum_{L=1}^\infty a^L 
M^\Ell (s,t; \eps)
&=&
\exp\left\{ \suml  a^\ell   
\left[ f^\pel (\eps) M^\One (s,t; \ell \eps) + h^\pel (s,t;\eps)\right]
\right\}
\label{bds}
\eea
where
\be
f^\pel (\eps)  
=
  {1\over 4} \gamma^\pel
+ {1\over 2} \eps \: \ell \: \cG_0^\pel 
+ \eps^2  f_2^\pel  
\label{deff}
\ee
with $\gamma^\pel$ and $\cG_0^\pel $ defined in \eqn{defanom},
and $h^\pel(s,t;\eps)$, which is finite as $\eps \to 0$,
contains information about the short-distance behavior of the amplitude.
The ratio of the one-loop amplitude to the tree amplitude is
\be
M^\One (s,t;\eps) = - \half st\, I_4^\One (s,t)
\label{defM}
\ee
where the scalar box integral
\be
I_4^\One (s,t)
= -i \mu^{2\ep} \e^{\ep \gamma} \pi^{-D/2}  \int 
{d^D p \over p^2 (p-k_1)^2 (p-k_1-k_2)^2 (p+k_4)^2 } 
\ee
may be evaluated exactly in terms of the 
hypergeometric function \cite{Bern:1993kr}.
The BDS conjecture (\ref{bds}) for the four-gluon amplitude 
is wholly consistent with the IR-divergence structure
as reviewed in sec.~\ref{secIR} and appendix \ref{appa}, 
but goes beyond it to assert that 
$h^\pel(s,t;\eps)$ is independent of $s$ and $t$ in the limit $\eps \to 0$. 

In the Regge limit $s \gg -t$, one finds
\cite{Bern:1998sc,DelDuca:2008pj},
neglecting terms suppressed by $\cO(t/s)$, 
\bea
M^\One (s,t;\eps)
&=&
\left( \mu^2 \over -t \right)^\eps  {r(\eps)\over \eps}
\left[  \log \left(- {s \over t} \right) -i \pi
+ \psi(1+\eps) - 2 \psi(-\eps) + \psi(1) 
\right] +  \cO \left( {t \upon  s} \right) 
\nonumber\\
&=& 
\left( \mu^2 \over -t \right)^\eps  \reps
\left[  
- { 2 \over \eps^2}
+ {1 \over \eps} \log \left(-{s \over t} \right)  
-{i \pi \over \eps}
+  \sum_{m=0}^\infty [ 2 + (-1)^m ] \zeta_{m+2}  \eps^m  
\right] +  \cO \left( {t \upon  s} \right) 
\nonumber
\\ 
\label{MstRegge}
\eea
where
\be
\reps 
= {\Gamma(1+\eps) \Gamma(1-\eps)^2 \over \Gamma(1-2\eps)} \e^{\gamma \eps}
= 1 - {1\over 2} \zeta_2  \eps^2
- {7 \over 3} \zeta_3  \eps^3 - {47 \over 16} \zeta_4  \eps^4 + \cdots
\label{defr}
\ee
If the $h^\pel(s,t;\eps)$ term were absent from \eqn{bds}, then
\eqn{MstRegge} would suffice to establish that 
$A^\Ellzero$  goes as $\log^L (-s/t)$ in the Regge limit,
as claimed in \eqn{planarconj}.
This claim would still be valid, even with the
$h^\pel(s,t;\eps)$ term present, provided that 
$h^\pel(s,t;\eps)$ grows no faster than 
$\log^\ell (-s/t)$ in the Regge 
limit.\footnote{The $h^\pel(s,t;\eps)$ 
can affect the coefficients of nonpositive powers of $\eps$ 
in $A^\Ellzero$ 
through interference with the IR-divergent 
terms in $M^\One(s,t; \ell\eps)$.}
In fact, the situation may be better than this.
Using the explicit expressions in ref.~\cite{Bern:2005iz}
together with the help of the Mathematica package HPL \cite{Maitre:2005uu}
we find that
\bea
h^\Two (s,t; \eps) 
& = &
   -\frac{\pi ^4}{72}
     +\left(-\frac{11 \pi ^4 }{360} 
              \left[ \log \left(- {s \over t} \right) - i \pi \right]
       -\frac{39 }{2} \zeta_5     
        +\frac{23 \pi^2}{12} \zeta_3\right) \eps
\label{defhtwo}
\\
&& +\left(
       \left[ \frac{41}{2} \zeta_5 +\frac{\pi^2}{4}  \zeta_3 \right] 
              \left[ \log \left(- {s \over t} \right) - i \pi \right]
         -15 \zeta_3^2-\frac{1789 \pi ^6}{30240}\right)
    \eps^2  +\cO(\eps^3) + \cO(t/s)
\nonumber
\eea
so that $ h^\Two (s,t; \eps) $ only grows as $\log(-s/t)$,
at least to $\cO(\eps^2)$.
If we make the assumption that 
$h^\pel(s,t;\eps)$ grows less strongly than  $\log^{\ell} (-s/t)$ in the
Regge limit for all $\ell$, 
then it would make no contribution to the 
leading log behavior of the planar $L$-loop amplitude, 
and we could conclude 
that\footnote{Interestingly, the individual scalar $L$-loop diagrams
that contribute to the planar $L$-loop amplitude generically
behave as \eqn{expectation} in the Regge limit, 
but all powers of $\log (-s/t)$ higher than $L$ cancel when they
are added up.}
\be
A^\Ellzero_\ii{1} 
\regge  
{1 \over L!}
\TREE
\left( \mu^2 \over -t \right)^{L\eps}  
\left(\reps \over \eps\right)^L
\left[ \log \left(-{s \over t }\right)  \right]^L
+ {\rm subleading}
\label{AEllRegge}
\ee
This behavior is precisely in accord with \eqn{planarconj},
with $( r(\eps)/\eps )^L$ yielding constants $c_{L+m}$
with the expected degree of transcendentality. 

Now we consider the Regge limits of the other color-ordered 
amplitudes\footnote{Also, recall that
$A^\Ellzero_\ii{4} = A^\Ellzero_\ii{3}$,
$A^\Ellzero_\ii{5} = A^\Ellzero_\ii{2}$,
and $A^\Ellzero_\ii{6} = A^\Ellzero_\ii{1}$.
}
\be
A^\Ellzero_\ii{2} = M^\Ell (s,u; \eps)  ~ A^\Zero_\ii{2},
\qquad
A^\Ellzero_\ii{3} = M^\Ell (t,u; \eps)  ~ A^\Zero_\ii{3}.
\label{other}
\ee
These are also given by the BDS ansatz.
To obtain $A^\Ellzero_\ii{3}$ we replace
$\log(-s/t)-i\pi$ with 
$\log(u/t) = \log(-s/t) + \cO(t/s)$
in \eqn{MstRegge} 
to obtain\footnote{Terms which are subleading in $t/s$ can 
in principle lead to subleading Regge trajectories and/or cuts,
which we do not examine in this paper.  The terms 
of $\cO(t/s)$ relative to the terms we keep could in principle
lead to Regge trajectories passing through $j=0$ at $t=0$.
This possibility is investigated in ref.~\cite{Grisaru:1982bi}.}
\be
\hspace{-1mm}
M^\One (t,u; \eps) 
= 
\left( \mu^2 \over -t \right)^\eps  \reps
\left[  - { 2 \over \eps^2}
+ {1 \over \eps} \log \left(-{s \over t} \right)  
+  \sum_{m=0}^\infty [ 2 + (-1)^m ] \zeta_{m+2}  \eps^m  
\right] +  \cO \left( {t \upon  s} \right) 
\label{MtuRegge}
\ee
Then, again subject to the assumption 
about $h^\pel(s,t;\eps)$ made above,
$A^\Ellzero_\ii{3}$ 
also has leading log behavior in the Regge limit
given by \eqn{AEllRegge}.

On the other hand,
$ M^\One (s,u; \eps) $ grows faster than $\log(-s/t)$ in the Regge limit,
\bea
M^\One (s,u; \eps) 
&=& 
\left( \mu^2 \over -t \right)^\eps 
\left[   - { 2 \over \eps^2} 
         + {2 \over \eps} \log \left(-{s \over t} \right) 
         - {i \pi \over \eps} 
	 - \log^2 \left(-{s \over t} \right) 
	+ i \pi \log \left(-{s \over t} \right) 
 	+ 4 \zeta_2
		+ \cO(\eps) \right]
\nonumber\\
&&\hspace{100mm} +  \cO \left( {t \upon  s} \right)  \,,
\label{MsuRegge}
\eea
and so $M^\Ell (s,u; \eps)$ grows faster than $\log^L(-s/t)$.
This apparent contradiction to \eqn{planarconj}
is resolved by recognizing that $A^\Ellzero_\ii{2}$ 
is suppressed by $t/s$ relative to
$A^\Ellzero_\ii{1}$ and $A^\Ellzero_\ii{3}$,
because $A^\Zero_\ii{2} = -4iK/su$,
and is therefore entirely contained in the ``subleading'' term.
In addition, the $ - \log^2(-s/t)$ dependence in \eqn{MsuRegge}
will lead to exponential suppression of the Regge trajectory
associated with this amplitude,
as we will see in sec.~\ref{secregge}.

\subsection{Regge limit of $A^\Oneone$} 
\label{seconelimit}

In this paper, we are particularly interested in 
the Regge behavior of subleading-color amplitudes.
The simplest case is the one-loop subleading-color amplitude, which
is given by \cite{Green:1982sw}
\be
A^\Oneone_\ii{7} =
A^\Oneone_\ii{8} =
A^\Oneone_\ii{9} 
=
2 \left( A^\Onezero_\ii{1} + A^\Onezero_\ii{2} + A^\Onezero_\ii{3}  \right).
\label{oneone}
\ee
We use eqs.~(\ref{bds}), (\ref{MstRegge}), (\ref{other}), and (\ref{MtuRegge}),
and recall that $A^\Onezero_\ii{2}$ is suppressed by $t/s$,
to obtain,
in the Regge limit,
\bea
A^\Oneone_\ii{7} 
&=&
\TREE
\left( \mu^2 \over -t \right)^\eps 
\left[  
-{2\pi i \reps\over \eps} + 
\cO \left( {t \upon  s} \right) 
\right]
\\
&=&
\TREE
\left( \mu^2 \over -t \right)^\eps 
\left[  
-{2\pi i \over \eps} 
+ i \pi \zeta_2  \eps
+ {14 \pi i \over 3} \zeta_3  \eps^2 
+ {47  \pi i \over 8} \zeta_4  \eps^3 + \cdots
+ \cO \left( {t \upon  s} \right) 
\right].
\nonumber
\eea
This confirms the conjectured behavior (\ref{nonplanarconj})
in the case $L=k=1$.
It was previously shown in eq.~(40) of ref.~\cite{DelDuca:1998kx}
that the real part of $A^\Oneone_\ii{7}$ vanishes
to $\cO (t/s)$.

\subsection{Regge limits of $A^\Twoone$ and $A^\Twotwo$} 
\label{sectwolimit}

The two-loop subleading-color amplitudes 
$\ket{A^\Twoone}$ and $\ket{A^\Twotwo}$
are known exactly \cite{Bern:1997nh}.
The former is given by
\bea
\hspace{-5mm}
A_\ii{7}^\Twoone &=& 
-2 i K
\Bigl[
s \left( 3 I_4^\TwoP(s, t) + 2 I_4^\TwoNP(s, t)
+ 3 I_4^\TwoP(s, u)  + 2 I_4^\TwoNP(s, u)\right)
\label{twoloopDT}
\\ && \hspace{11mm}
-t \left(I_4^\TwoNP(t, s) + I_4^\TwoNP(t, u)\right)
-u \left(I_4^\TwoNP(u, s) + I_4^\TwoNP(u,t)\right)
\Bigr]  
\nonumber
\eea
where $A^\Twoone_\ii{8}$ and $A^\Twoone_\ii{9}$ 
may be obtained via cyclic permutations of $s$, $t$, and $u$.
The two-loop  planar and non-planar scalar integrals appearing in \eqn{twoloopDT}
are
\bea
I_4^\TwoP(s,t) 
&=&
\left( -i \mu^{2\ep} \e^{\ep \gamma} \pi^{-D/2} \right)^2
\int 
 {d^D p \, d^D q
\over p^2 \, (p + q)^2 q^2 \, (p - k_1)^2 \,(p - k_1 - k_2)^2 \,
        (q-k_4)^2 \, (q - k_3 - k_4)^2 } 
\nonumber\\
I_4^\TwoNP(s,t)  
&= &
\left( -i \mu^{2\ep} \e^{\ep \gamma} \pi^{-D/2} \right)^2 \int 
{d^D p \, d^D q\over p^2\,(p+q)^2\, q^2 \, (p-k_2)^2  \,(p+q+k_1)^2\,
  (q-k_3)^2 \, (q-k_3-k_4)^2}   \,.
\nonumber\\
\eea
Explicit expressions for these integrals
are given in refs.~\cite{Bern:2005iz}  and \cite{Tausk:1999vh} respectively.
Continuing these expressions to the physical region $s>0$, $t,u<0$,
and dropping terms suppressed by $t/s$ in the Regge limit, we obtain
\bea
A^{(2,1)}_{[7]} 
= 
\TREE
\left(\mu^2 \over -t\right)^{2\epsilon}  
&\Bigg\{& 
 \frac{4 i \pi  }{\epsilon ^3}
-\frac{3 i \pi  \Bigl[ \log (\Y) -i \pi\Bigr]  }{\epsilon ^2}
-\frac{3 i \pi ^3 }{2 \epsilon } 
\nonumber\\&&
+\frac{i \pi}{6} \left[3 \pi^2 \log (\Y) -82 \zeta_3 -3 i \pi^3 \right]
+\cO\left(\epsilon\right)
+\cO \left( {t \upon  s} \right) 
\Bigg\}
\nonumber
\\
A^{(2,1)}_{[8]} 
= \TREE
\left(\mu^2 \over -t\right)^{2\epsilon}  
&\Bigg\{& 
 \frac{4 i \pi }{\epsilon ^3}
-\frac{3 i \pi  \log (\Y)}{\epsilon ^2}
-\frac{3 i \pi ^3}{2 \epsilon }
\nonumber\\ &&
+\frac{i \pi}{6} \left[3 \pi^2 \log (\Y)-82 \zeta_3\right]
+\cO\left(\epsilon \right)
+\cO \left( {t \upon  s} \right) 
\Bigg\}
\nonumber\\ A^{(2,1)}_{[9]} 
= \TREE
 \left(\mu^2 \over -t\right)^{2\epsilon}  &\Bigg\{& 
    \frac{4 i \pi }{\epsilon ^3}
- \frac{3 i \pi ^3}{2 \epsilon }
-\frac{95}{3} i \pi  \zeta_3
+\cO\left(\epsilon \right)
+\cO \left( {t \upon  s} \right) 
\Bigg\}.
\label{twoloopone}
\eea
By using eq.~(\ref{XYZlimit}),
one may easily verify that that IR-divergent parts of this expression
agree with the general expression (\ref{tedious})
derived in the last section.

The most-subleading-color two-loop amplitudes are given by \cite{Bern:1997nh}.
\bea
\hspace{-9mm}
A_\ii{1}^\Twotwo &=& 
-2 i K \,
\Bigl[
s \left(I_4^\TwoP(s, t) + I_4^\TwoNP(s, t)
+ I_4^\TwoP(s, u)  + I_4^\TwoNP(s, u)\right)
\label{twoloopSC}
\nonumber\\ && \hspace{11mm}
+t \left(I_4^\TwoP(t, s) + I_4^\TwoNP(t, s)
+ I_4^\TwoP(t, u)  + I_4^\TwoNP(t, u)\right)
\\ && \hspace{10mm}
-2u \left(I_4^\TwoP(u, s) + I_4^\TwoNP(u, s)
+ I_4^\TwoP(u,t)  + I_4^\TwoNP(u,t)\right)
\Bigr]  .
\nonumber
\eea
The other single-trace amplitudes $A^\Twotwo_\ii{i}$ 
are obtained by making the appropriate permutations of $s$, $t$, and $u$ 
in this expression.
Again extracting the Regge limit of these amplitudes, we find
\bea
A^{(2,2)}_{[1]} 
=A^{(2,2)}_{[6]} 
= \TREE
\left(\mu^2 \over -t\right)^{2\epsilon}
&\Bigg\{ &
  \frac{ i \pi   \Bigl[ \log (\Y)  + i \pi\Bigr]}{\epsilon ^2}
-\frac{i \pi }{6} 
\left[ \pi ^2 \log (\Y) +i \pi ^3 + 36 \zeta_3  \right]
\nonumber\\&&
+\cO \left(\epsilon\right)
+\cO \left( {t \upon  s} \right) 
\Bigg\}
\nonumber\\[2mm]
A^{(2,2)}_{[2]}
=A^{(2,2)}_{[5]}
= \TREE
 \left(\mu^2 \over -t\right)^{2\epsilon} &\Bigg\{ &
  \frac{ i \pi \Bigl[- 2 \log (\Y) + i \pi\Bigr] }{\epsilon^2}
+ \frac{i \pi }{6} \left[
 2 \pi^2 \log (\Y) -i \pi^3 +72 \zeta_3  \right]
\nonumber\\&&
+\cO \left(\epsilon\right)
+\cO \left( {t \upon  s} \right) 
\Bigg\}
\nonumber\\[2mm]
A^{(2,2)}_{[3]}
=A^{(2,2)}_{[4]}
= \TREE
 \left(\mu^2 \over -t\right)^{2\epsilon}& \Bigg\{ &
\frac{ i \pi  \Bigl[\log (\Y) - 2 \pi i\Bigr] }{\epsilon ^2}
-\frac{i \pi }{6} 
\left[  \pi^2  \log (\Y) -2 i \pi^3   +36 \zeta_3  \right]
\nonumber\\&&
+\cO \left(\epsilon \right)
+\cO \left( {t \upon  s} \right) 
\Bigg\}.
\label{twolooptwo}
\eea

We see that both subleading-color amplitudes
$\ket{A^\Twoone}$ and $\ket{A^\Twotwo}$
go as $\log(-s/t)$ in the Regge limit (at least through $\cO(\eps^0)$),
thus adding support to our conjecture (\ref{nonplanarconj}).

\subsection{Regge limit of IR-divergences of higher-loop amplitudes}
\label{secrecurse}

In sec.~\ref{seconelimit},
we saw that the one-loop subleading-color amplitude goes as 
$\log^0(-s/t) $
to all orders in $\eps$,
and in sec.~\ref{sectwolimit}
that all two-loop subleading-color amplitudes go as 
$\log^1(-s/t)$, at least through $\cO(\eps^0)$.
Thus suggests that,
while the $L$-loop {\it planar} amplitude 
(probably) goes as $\log^L (-s/t)$ in the Regge limit,
the $L$-loop {\it subleading-color} amplitudes 
only go as $\log^{L-1} (-s/t)$ in the Regge limit, 
as conjectured in \eqn{nonplanarconj}.

Because the IR-finite parts of the subleading-color amplitudes 
beyond two loops are not known explicitly,
we cannot prove this conjecture, 
but in this section we will perform an important consistency check.   
We will prove that the {\it IR-divergent}
contributions to the $L$-loop subleading-color amplitudes
grow no faster than $\log^{L-1}(-s/t)$
in the Regge limit, 
provided that $\Lp$-loop subleading-color amplitudes 
(both IR-divergent and finite parts)
grow no faster than $\log^{\Lp-1}(-s/t)$ 
for all $\Lp<L$.
Thus, with this inductive argument, 
it is sufficient to prove that the 
{\it IR-finite} contribution to the $L$-loop subleading-color 
amplitudes goes as $\log^{L-1}(-s/t)$ 
to establish it for the full amplitude.

Our first step is to prove a weaker result,
namely that
the IR-divergent part of any $L$-loop amplitude
(planar or subleading) 
grows no faster than $\log^L(-s/t)$,
provided that no $\Lp$-loop amplitude 
(planar or subleading) 
with $\Lp < L$ grows faster than $\log^\Lp(-s/t)$.
Consider $\bG^\pel/N^\ell$ defined by \eqns{newG}{defGamOne}, 
with $Q^2 = -t$,
and $\ia$ through $\id$ given by \eqn{defnewalpha}.
In \eqn{alphalimit}, we show that
$\ia$ through $\id$,
and therefore $\bG^\pel/N^\ell$,
go as $\log(-s/t)$ in the Regge limit.
Consequently, the strongest growth of 
any (IR-divergent) term in \eqn{expandinit}
is $\log^q(-s/t)$
where $q =\ell_0+ \suml  n_\ell$.
Since $q \le L$ by \eqn{defLk}, we have established our result.

Now we prove a stronger result, 
namely that the $\log^L(-s/t)$ terms 
are actually absent from the IR-divergent
contributions to $L$-loop {\it subleading-color} amplitudes.
The only terms in \eqn{expand} 
that could yield $\log^L(-s/t)$ growth are those 
with $n_\ell = 0$ for $\ell > 1$
(so that the inequality $q \le L$ is saturated)   
and containing $\ket{H^\Lpzero}$ 
(since we assume that $\ell_0$-loop subleading-color
amplitudes grow no faster than $\log^{\ell_0-1}(-s/t)$), 
namely, terms of the form
\be
\left( \gz + {1 \over N} \gone \right)^{L-\ell_0}  \ket{H^\Lpzero(\eps)}.
\ee
First we consider the $\ell_0=0$ term 
\be
\left( \gz + {1 \over N} \gone \right)^{L}  \ket{\Azero}
\ee
with $\ket{\Azero}$ given in the Regge limit by
\be
\ket{A^\Zero} \regge
- \frac{ 4 i K }{st} \left(1,0,-1,-1,0,1,0,0,0\right)^T .
\label{reggetree}
\ee 
Since any subleading-color amplitude contains at least one
factor of $\gone$, we can see that
the structure $\cdots \ic \ia^n \ket{\Azero}$ for some $n\ge 0$
will always appear. 
By virtue of \eqns{reggetree}{alphalimit},
one can see that the leading log term in 
$\cdots \ic \ia^n \ket{\Azero}$ vanishes 
since $\ia$ doesn't change the structure of $\ket{A^\Zero}$ 
and $\ic$ annihilates it.

Essentially the same argument works for the
$\ell_0 \neq 0$ terms as well.
By the BDS ansatz, the leading-color amplitudes  $\ket{A^\Lpzero (\eps)}$
are given by 
\be
A^\Lpzero_\ii{1} = M^{(\Lp)} (s,t)  A^\Zero_\ii{1}, \qquad
A^\Lpzero_\ii{3} = M^{(\Lp)} (u,t)  A^\Zero_\ii{3}.
\ee
But the leading log terms of
$ M^{(\Lp)} (u,t)$ and $M^{(\Lp)} (s,t) $ are equal 
in the Regge limit,
so the leading log piece of 
$\ket{A^\Lpzero (\eps)}$
(and therefore of the IR-finite contribution $\ket{H^\Lpzero (\eps)}$)
is proportional to $\ket{\Azero}$.
Thus the putative $\log^L(-s/t)$ terms of $ \cdots \ic \ia^n \ket{H^\Lpzero}$ 
also vanish.

Hence, we conclude that the IR-divergent terms of the
$L$-loop subleading-color amplitudes go as $\log^{L-1}(-s/t)$,
provided that the same holds for all lower-loop subleading-color
amplitudes.

\section{Regge trajectories}
\setcounter{equation}{0}
\label{secregge}

In sec.~\ref{seclimit},
we discussed the leading log behavior of the $L$-loop
planar and subleading-color amplitudes 
in the Regge limit.  
In this section, we will sum the loop amplitudes 
to obtain Regge trajectories.

\subsection{Planar gluon Regge trajectory} 
\label{secplanartraj} 

We first focus on the planar color-ordered amplitude 
$A_\ii{3}$,
which is real in the region $s>0$,  $t,u<0$,
and whose Regge behavior was explored in 
refs.~\cite{Drummond:2007aua, Naculich:2007ub,DelDuca:2008pj}.
The $\log^L(-s/t)$ behavior of the Regge limit
of the planar $L$-loop amplitude
$A^{\Ellzero}_\ii{3}$ 
conjectured in sec.~\ref{seclimit}
suggests that the all-orders planar amplitude exhibits Regge 
behavior
\be
\sumL a^L 
A^{\Ellzero}_\ii{3} 
\regge
\beta_0 (t) \left( - {s \over t} \right)^{\alpha_0(t)}
\label{reggetraj}
\ee
where  $\alpha_0(t)$ is the 
Regge trajectory function,
and $\beta_0 (t)$ the Regge residue.
Indeed, using eqs.~(\ref{bds}), (\ref{other}), and (\ref{MtuRegge}),
one obtains the following expression
for the Regge trajectory function \cite{DelDuca:2008pj}
\be
\alpha_0(t) = 1 + \suml f^\pel (\eps) 
{\relleps \over \ell \eps} 
a^\ell \left( \mu^2 \over -t \right)^{\ell\eps}
+ \cO(\eps)\,.
\label{firstalphanought}
\ee
where $\cO (\eps)$ corrections come from the 
$h^\pel(s,t;\eps)$ terms\footnote{
In fact, Regge behavior (\ref{reggetraj}) will hold to 
all orders in $\eps$ only if $h^\pel(s,t;\eps)$ grows
no faster than $\log(-s/t)$ for all $\ell$.} 
in \eqn{bds}.
For example, the $\log(-s/t)$ dependent terms in \eqn{defhtwo}
contribute to the two-loop Regge trajectory \cite{DelDuca:2008pj}
at $\cO(\eps)$ and $\cO(\eps^2)$.
The leading 1 comes from the tree amplitude
$ A^\Zero_\ii{3} = -4iK/ut$ 
since 
\be
\label{defk}
-4iK \regge k s^2
\ee
where $k$ depends on the helicities of the gluons,
and is finite as $s \to \infty$.
We can rewrite 
\eqn{firstalphanought} as \cite{Drummond:2007aua, Naculich:2007ub}
\bea
\alpha_0(t) 
&=& 
1 + {1\over 4 \eps} \suml   a^\ell {\gamma^\pel \over \ell} 
       \left( \mu^2 \over -t \right)^{\ell\eps}
  + {1\over 2} \suml  a^\ell  {\cG_0^\pel} 
	+ \cO(\eps) 
\nonumber\\
&=& 
1 + {1\over 4 \eps} \gamma^{(-1)} (a)  
  - {1\over 4} \gamma (a) \log\left(-t\over\mu^2\right)
  + {1\over 2} \cG_0(a)
	+ \cO(\eps) .
\label{defalphanought}
\eea
The residue is given by \cite{Naculich:2007ub}
\be
\beta_0 (t) 
=
k
\exp\Bigg\{ 
-\frac{1}{2\eps^2} 
\gamma^{(-2)} 
\left( \mu^{2\eps} a \over (-t)^\eps  \right)
-\frac{1}{\eps} 
\cG_0^{(-1)} 
\left( \mu^{2\eps} a \over (-t)^\eps  \right)
+ \zeta_2 \gamma(a) 
- 2 f_2^{(-2)}  (a)
+ h(a) 
+ \cO(\eps)
\Bigg\}
\label{defbetanought}
\ee
where the functions in \eqns{defalphanought}{defbetanought}
are defined by
\bea
&&
\gamma^{(-1)}(a) = \suml {a^\ell \over \ell} \gamma^\pel, \qquad
\gamma^{(-2)}(a) = \suml {a^\ell \over \ell^2} \gamma^\pel, \qquad
\\
&&
\cG_0^{(-1)} (a) = \suml {a^\ell \over \ell}  {\cG_0^\pel}, \qquad
f_2^{(-2)}(a) = \suml {a^\ell\over \ell^2}  {f_2^\pel},\qquad
h(a) = \suml a^\ell h^\pel(0), \quad
\nonumber
\eea
Explicitly, we have \cite{Bern:2005iz}
\be
\zeta_2 \gamma(a) - 2 f_2^{(-2)}  (a) + h(a) 
= 4 \zeta_2 a - \frac{43}{4} \zeta_4 a^2 + 
\left( \frac{8657}{216} \zeta_6 - \frac{17}{9} \zeta_3^2 \right) a^3 + \cdots
\ee
and 
$\gamma^\pel$ and $\cG_0^\pel$ are given in \eqn{defanom}.

We now consider the other color-ordered amplitudes 
$A_\ii{1}$ and $A_\ii{2}$.
Using \eqn{MstRegge}, 
one can see that the planar contribution to $A_\ii{1}$ 
goes in the Regge limit to 
\be
\sumL a^L 
A^{\Ellzero}_\ii{1} 
\regge
\beta_0 (t) \e^{-i \pi \alpha_0(t)} \left( - {s \over t} \right)^{\alpha_0(t)}.
\label{othertraj}
\ee
The presence of the $-\log^2 (-s/t)$ term in \eqn{MsuRegge},
however, results in the exponential suppression of $A_\ii{2}$
in the Regge limit, \viz, $(-s/t)^{- \log(-s/t) } \to 0$.

We now rewrite the full planar four-gluon amplitude (\ref{fourgluon}) as 
\bea
\cA^{\rm planar}_{4-{\rm gluon}}
&=&
g^2 \sumL a^L \left[ 
\left( A^\Ellzero_\ii{1} - A^\Ellzero_\ii{3} \right) \ff
\right.
\label{fullplanar}
\\
&&\hspace{14mm}
\left. +
\left( A^\Ellzero_\ii{1} + A^\Ellzero_\ii{3} \right) \dd 
+
A^\Ellzero_\ii{2} \left( \cC_\ii{2} + \cC_\ii{5} \right) 
\right]
\nonumber
\eea
where
\be
\ff 
=
\half \left( \cC_\ii{1} -\cC_\ii{3} -\cC_\ii{4} + \cC_\ii{6}    \right)
,\qquad
\dd
=
\half\left( \cC_\ii{1} +\cC_\ii{3} +\cC_\ii{4} + \cC_\ii{6}   \right).
\ee
The coefficient of the $\ff$ term in \eqn{fullplanar} 
corresponds to the exchange of a trajectory 
in the $t$ channel with the quantum numbers of the gluon,
and so the planar gluon Regge trajectory is given by
\be
\sumL a^L \left( A^\Ellzero_\ii{1} - A^\Ellzero_\ii{3} \right)
\regge
B_0(t)
\left( - {s \over t} \right)^{\alpha_0(t)}
\label{gluontraj}
\ee
where $\alpha_0(t)$,  given in \eqn{defalphanought},
represents the planar gluon Regge trajectory, and 
$B_0(t)$, given by $\beta_0 (t) \left( \e^{-i \pi \alpha_0(t)} -1  \right)$,
is the Regge residue, including the signature factor.
The coefficient of $\dd$ in \eqn{fullplanar} 
gives a wrong signature trajectory,
and the $A_\ii{2}$ term is exponentially damped in the Regge limit.

\subsection{$1/N^2$ corrections} 
\label{seccorrtraj}

As seen in the previous section, the planar amplitudes sum up to give the
planar gluon Regge trajectory (\ref{gluontraj}). 
It might be expected that the full amplitude
would give rise to subleading-color corrections to the gluon trajectory.
Let us characterize the first subleading-color corrections
to the gluon trajectory as
\be
A_\ii{1} - A_\ii{3} 
\regge
\left[ B_0 (t) + {1\over N^2} B_2(t) + \cdots \right]
\left( - {s \over t} \right)^{\alpha_0(t) + (1/N^2) \alpha_2(t)   + \cdots}
\label{corrtraj}
\ee
The $1/N^2$ corrections to the amplitude may be (at least partially) summed
to give
\bea
&&\hspace{-12mm}
\sum_{L=2}^\infty 
{a^L  \over N^{2} }
A^\Elltwo_\ii{1}
\to
\exp \left[ - {2 a \over \eps^2} + \cdots \right]
{ a^2 \over N^2 }
\left(\mu^2 \over -t\right)^{2\eps} {i \pi  k  s \over t }  
\Bigg(
  \frac{ \log (\Y)  + i \pi}{\epsilon ^2}
-\frac{\pi^2}{6} 
\log (\Y) -{i \pi ^3\over 6} -  6 \zeta_3  
\Bigg)
\nonumber\\
&&\hspace{-12mm}
\sum_{L=2}^\infty 
{a^L  \over N^{2} }
A^\Elltwo_\ii{3}
\to
\exp \left[ - {2 a \over \eps^2} + \cdots \right]
{ a^2 \over N^2 }
\left(\mu^2 \over -t\right)^{2\eps} { i \pi  k  s \over t }  
 \Bigg(
\frac{ \log (\Y) - 2 \pi i}{\epsilon ^2}
-\frac{\pi^2}{6} 
\log (\Y) + {i \pi^3\over3}  -  6 \zeta_3 
\Bigg)
\nonumber\\
\eea
where the IR-finite terms are obtained from the two-loop
subleading-color amplitude (\ref{twolooptwo}),
and the exponential prefactor results from
summing the leading IR-divergent term (\ref{even}) to all orders in $L$.
All the
$\log(-s/t)$ terms cancel from the combination
of amplitudes that contributes to the gluon Regge trajectory
\be
\sum_{L=2}^\infty 
{a^L  \over N^{2} }
\left( A^\Elltwo_\ii{1}
     - A^\Elltwo_\ii{3} \right)
\to 
{ a^2 \over N^2 }
\left(\mu^2 \over -t\right)^{2\eps} {  k  s \over t }  
 \Bigg(
- \frac{  3 \pi^2}{\epsilon ^2}
+ {\pi^4\over2} 
\Bigg)
+\cO(a^3)
\label{residue}
\ee
and consequently, the gluon Regge trajectory function  $\alpha_0(t)$
remains uncorrected through $\cO(a^2)$,
as might be anticipated from the corresponding 
two-loop result for QCD \cite{Fadin:1996tb,DelDuca:2001gu}.
The expression (\ref{residue}) corresponds to a $1/N^2$ correction 
\be
B_2(t) = 
k a^2 
\left(\mu^2 \over -t\right)^{2\eps} 
 \Bigg(
\frac{  3 \pi^2}{\epsilon ^2}
- {\pi^4\over2} 
\Bigg) + \cO(a^3)
\ee
to the Regge residue starting at two loops \cite{DelDuca:2001gu}.

\subsection{Regge trajectory for double-trace amplitudes}
\label{secdoubletraj}

In sec.~\ref{seclimit},
we presented evidence that the $L$-loop subleading-color amplitudes
go as $\log^{L-1} (-s/t)$ in the Regge limit.
This suggests that the double-trace amplitudes may
also exhibit Regge behavior 
\be
\sum_{L=1}^\infty a^{L-1} \ket{A^\Ellone}  
\regge
\beta_1(t) \left( - {s \over t} \right)^{\alpha_1(t)}.
\ee
We will now see how far this expectation is borne out.

In sec.~\ref{secIREllone}, 
we calculated the first three IR-divergent terms
of the subleading-color amplitude $\ket{A^\Ellone}$.
For the moment, let us focus on only one component
\bea
A_\ii{8}^\Ellone 
&=& 
\TREE
\frac{Y \eps}{(L-1)!}  
\left[ \frac{-2}{\eps^2} \left(\mu^2 \over -t\right)^{\epsilon}   \right]^{L}
\Bigg\{  
1
+ \frac{3}{4}  (L-1)  X \eps
\nonumber\\
&&
+ \frac{7}{24} (L-1)(L-2) X^2 \eps^2 
+ \frac{1}{8} (L^2 - 17 L + 12) \zeta_2 \eps^2
+ \cO (\eps^3)  
+ \cO \left( t \upon s \right) 
\Bigg\}
\nonumber\\ 
&=&
\TREE
{-2 Y \over \eps} 
\left(\mu^2 \over -t\right)^{\epsilon}  
\frac{1}{(L-1)!}  
\left[\left(\mu^2 \over -t\right)^{\epsilon}  
\left(- \frac{2}{\eps^2} - \frac{3 X}{2 \eps} \right) \right]^{L-1}
\label{eight}\\
&&
\times \Bigg\{ 1 + \frac{1}{96} (L-1)(L-2) X^2 \eps^2 
+ \frac{1}{8} (L^2 -17L + 12) \zeta_2 \eps^2
+ \cO (\eps^3) 
+ \cO \left( t\upon s \right) 
\Bigg\}.
\nonumber
\eea
Since, by \eqn{XYZlimit}, $ X^2 \gg 1 $ in the Regge limit,
we can drop the $\zeta_2$-dependent term in the curly braces in \eqn{eight}.
The series can be summed to obtain
\bea
\sum_{L=1}^\infty  a^{L-1} A_\ii{8}^\Ellone 
&=&
\TREE
{-2 Y \over \eps} 
\left(\mu^2 \over -t\right)^{\epsilon}  
\exp \left[
\left(\mu^2 \over -t\right)^{\epsilon}   
\left( -\frac{2 a}{\eps^2} - \frac{3 a X}{2 \eps} \right) 
\right]
\left[ 1 + \frac{a^2 X^2}{24 \eps^2} + \cdots \right]
\nonumber\\
&=&
\TREE
{-2 Y \over \eps} 
\left(\mu^2 \over -t\right)^{\epsilon}  
\exp \left[
\left(\mu^2 \over -t\right)^{\epsilon}   
\left( -\frac{2 a}{\eps^2} - \frac{3 a X}{2 \eps} \right) 
 + \frac{a^2 X^2}{24 \eps^2} + \cdots \right]
\nonumber\\
&=&
{2 \pi i k  \over \eps} 
\left(\mu^2 \over -t\right)^{\epsilon}  
\exp \left[- \frac{2a }{\eps^2} \left({\mu^2 \over -t}\right)^{\epsilon}\right]
\left( s \over -t \right)^{ \alpha_1(t)
~+~ ({a^2}/{24 \eps^2}) \log (-s/t)  
}
\label{sumeight}
\eea
where in the last line of \eqn{sumeight}
we used \eqns{defk}{XYZlimit}.
The Regge trajectory function in \eqn{sumeight} is given by 
\be
\alpha_1(t)
=
1 + \frac{3a }{2 \eps} \left(\mu^2 \over -t\right)^{\epsilon}   
=
 1 + \frac{3a }{2 \eps} 
- \frac{3a }{2} \log\left( -t \over \mu^2 \right) 
+ \cdots 
\label{trajectory}
\ee
Equation (\ref{trajectory}) suggests a massless spin-1 state
with Regge slope 3/2 that of the planar (gluon) trajectory.
However, since \eqn{sumeight} cannot lead to a physical massless particle, 
we speculate that this is a trajectory 
which is nonsense-choosing\footnote{
See sec.~5 of ref.~\cite{Grisaru:1982bi} for a discussion
of possible nonsense-choosing states in $\cN=4$ SYM with gauge
group SU(2). In that reference, trajectories with possible massless
scalar bound states are also discussed, but not considered here,
as these are $\cO(t/s)$, and suppressed in the limits we consider. }
at $j=1$.
By contrast, the gluon lies on a trajectory
which chooses sense at $j=1$.
The $a^2 \log(-s/t)$ term in the exponent in \eqn{sumeight} 
can be interpreted as a Regge cut.

Starting from \eqn{tedious},
we obtain similar results in the Regge limit for $A^\Ellone_\ii{7}$:
\bea
\sum_{L=1}^\infty  a^{L-1} A_\ii{7}^\Ellone 
&=&
{2\pi i k \over \eps} 
\left(\mu^2 \over -t\right)^{\epsilon}  
\exp \left[
\left({\mu^2 \over -t}\right)^{\epsilon}
\left( - \frac{2a }{\eps^2} - \frac{3\pi i a}{2\eps} \right)
- \frac{\pi^2 a^2}{24 \eps^2}
\right]
\nonumber\\
&& \times
\left( s \over - t \right)^{  \alpha_1(t) - (i \pi a^2/12 \eps^2)
~+~ (a^2/24 \eps^2) \log (-s/t)  
}
\label{sumseven}
\eea
while 
\be
\sum_{L=1}^\infty  a^{L-1} A_\ii{9}^\Ellone 
=
{2\pi i k \over \eps} 
\left(\mu^2 \over -t\right)^{\epsilon}  
\exp \left[- \frac{2a }{\eps^2} \left({\mu^2 \over -t}\right)^{\epsilon}\right]
\left( s \over -t \right)^{1 \,  +\,   (i \pi a^2/6\eps^2) ~-~ ({a^2}/{6 \eps^2}) \log (-s/t)  
}
\label{sumnine}
\ee
has a fixed pole together with a Regge cut,
which leads to exponential damping.

\section{Conclusions}
\setcounter{equation}{0}
\label{secconcl}

Beginning with the assumption that all soft
anomalous dimension matrices $\bGam^\pel$ are proportional 
to $\bGam^\One$, and therefore commute with each other,
we derived all-loop-order expressions for the IR-divergent parts of
the planar and all subleading-color contributions
to the $\cN=4$ SYM four-gluon amplitude.
Explicit expressions for the leading IR divergences are presented
in \eqns{odd}{even},
confirming a conjecture of ref.~\cite{Naculich:2008ys}.
The first two terms in the Laurent expansion in the IR regulator $\eps$
are presented for the most-subleading-color amplitude $A^\EllL$
in \eqn{mostsubleading},
also confirming a conjecture of ref.~\cite{Naculich:2008ys}.
The three leading terms in the Laurent expansion in $\eps$
for $A^\Ellone$ are given in  \eqn{firstsubleading},
and their Regge limit in \eqn{tedious};  
further terms in the Laurent expansion could be computed as needed.

The iterative structure of planar amplitudes was exploited 
in ref.~\cite{Bern:2005iz} to formulate the BDS conjecture. 
No analogous results are known for subleading-color amplitudes.
A weaker possibility is that the amplitude
obtained by summing subleading-color amplitudes over all loops
has Regge behavior in the limit $s\to\infty$, $t$ fixed.
(It is weaker because $\cO(t/s)$ terms are neglected in this limit.
In contrast, the planar four-gluon amplitude is Regge 
exact \cite{Drummond:2007aua};
\ie, Regge behavior is manifest without taking any limit.) 
We first considered the Regge limit of four-gluon amplitudes,
and presented evidence that the leading logarithmic growth
of the subleading-color $L$-loop amplitudes is less severe than that
of the planar amplitudes, 
going as $\log^{L-1}(-s/t)$ 
rather than $\log^{L}(-s/t)$.
We then investigated $1/N^2$ corrections to the gluon 
Regge trajectory as well as Regge behavior of the subleading-color
double-trace amplitudes by summing over the IR-divergent 
parts of the $L$-loop amplitudes, neglecting terms of $\cO(t/s)$.
The subleading-color double-trace amplitudes exhibit Regge behavior:
that is, there is a Regge trajectory as well as a Regge cut which
emerges at three loops.  
Thus, in the weaker sense described in this paper, there is 
sufficient iterative structure to produce leading
Regge behavior in the subleading-color amplitudes.

\vspace{.2in}
{\bf Acknowledgments} 
The authors are grateful to Lance Dixon for correspondence and discussions,
to Johannes Henn for discussions,
and to Horatiu Nastase for his collaboration on 
refs.~\cite{Naculich:2008ew,Naculich:2008ys}.
We also thank Lance Dixon for drawing our attention to an error in
sec.~\ref{seccorrtraj} in {\tt v1} of this paper.

\vfil\break
\appendix
\section{Generalized ABDK equation}
\label{appa}
\setcounter{equation}{0}
\def\theequation{A.\arabic{equation}}

In this appendix we show that the IR-divergent part 
of the $L$-loop generalization \cite{Bern:2005iz}
of the ABDK relation \cite{Anastasiou:2003kj}
for the planar four-gluon amplitude
may easily be obtained from the expression (\ref{compact}) for
the four-gluon amplitude
\be
\sumL a^L \ket{A^\Ell (\eps) } 
=  \exp\left[ \suml {a^\ell \over N^\ell} \bG^\pel (\ell \eps) \right] 
\left( \sumL a^L \ket{H^\Ell (\eps) } \right)
\label{compactagain}
\ee

Consider the planar (leading-color) $L$-loop amplitude 
$\ket{A^\Ellzero}$, and its IR-finite part $\ket{H^\Ellzero}$,
which are proportional to the tree-level amplitude:
\be
A_\ii{1}^\Ellzero (\eps)= M^\Ell (\eps) A_\ii{1}^\Zero, \qquad
H_\ii{1}^\Ellzero (\eps)= \tM^\Ellf (\eps) A_\ii{1}^\Zero
\ee
{}From the expressions  
(\ref{newG}), (\ref{defGamOne}), and (\ref{defalpha}),
we observe that the leading-color term of $\bG^\pel$ 
is a diagonal matrix,
and moreover that all subleading corrections are off-diagonal.
Thus, retaining only the leading-color terms of \eqn{compactagain}, 
we have
\be
1 + \suml  a^\ell M^\pel (\eps)=
\exp\left[ \suml \frac{a^\ell}{N^\ell}  G_{[11]}^\pel (\ell \eps) \right]
\left( 1 + \suml  a^\ell \tM^\pelf (\eps)\right)
\ee
where $G^\pel_{[11]}$ denotes the $11$ matrix element of $\bG^\pel$.
Using \eqn{defX}, we may rewrite this as
\be
M^\pel(\eps) - X^\pel[M]  =
\frac{G_{[11]}^\pel (\ell \eps) }{N^\ell}  + \tM^\pelf(\eps) - X^\pel[\tM^{(f)}]  
\ee
which is valid to all orders in the $\eps$ expansion. 
Using \eqn{newG} we observe that
\bea
\frac{G_{[11]}^\pel (\ell \ep)} {N^\ell}
 &=& 
  {1 \over 2} \muQell
\left[ - \frac{\gamma^\pel}{(\ell \ep)^2}
       -\frac{2\cG_0^\pel} {\ell \ep} 
       +\frac{\gamma^\pel}{4 \ell \ep }\Gamma_{[11]}^\One\right]
\nonumber\\
&=&
\left[ \frac{\gamma^\pel}{4}  + \frac{\ell}{2} \cG_0^\pel \ep \right]
\muQell
\left[ - \frac{2}{(\ell \eps)^2} + 
\frac{\Gamma_{[11]}^\One}{2 \ell \eps}  \right] + \cO(\eps^0)
\nonumber\\
&=& 
f^\pel(\eps) 
\frac{G_{[11]}^\One  (\ell \ep)} {N}
+ \cO(\eps^0) 
\nonumber\\
&=&
f^\pel(\eps) M^\One (\ell \eps)  + \cO(\eps^0) 
\eea
where $f^\pel(\eps)$ is defined in \eqn{deff}.
Hence we obtain
\be
M^\pel(\eps) = X^\pel[M]  +
f^\pel(\eps) M^\One (\ell \eps)  + \cO(\eps^0) 
\ee
which is precisely the IR-divergent part of the generalized
ABDK relation for the four-gluon amplitude,
eq.~(4.13) of ref.~\cite{Bern:2005iz}.

\section{Explicit expressions for the four-gluon amplitude}
\label{appb}
\setcounter{equation}{0}
\def\theequation{B.\arabic{equation}}

In this appendix we collect various explicit expressions
for four-gluon amplitudes  needed in the paper.

The tree-level amplitudes are 
\be
\label{tree}
\ket{A^\Zero} = 
- \frac{ 4 i K }{stu}
\left(u, t, s, s, t, u, 0,0,0\right)^T
\ee 
where
$s$, $t$, and $u$ are the 
Mandelstam invariants  $s_{12} $, $s_{14}$, and $s_{13}$,
where $s_{ij} = (k_i + k_j)^2$,
with $s+t+u=0$ for massless external gluons.
The factor $K$, defined in eq.~(7.4.42) of ref.~\cite{Green:1987sp},
depends on the momenta and helicity of the external gluons, 
and is totally symmetric under permutations of the external legs.

The one-loop soft anomalous dimension matrix is given 
by \cite{MertAybat:2006mz}
\be
\label{oneloopanom}
\bGam^\One
=
- \frac{1}{N}  \sum_{i=1}^4 \sum_{j\neq i}^4 \bT_i \cdot \bT_j
 \log \left( {-s_{ij} \over  Q^2 } \right) 
\ee
where $\bT_i \cdot \bT_j = T_i^a T_j^a$
with $T_i^a$ the SU$(N)$ generators in the adjoint representation.
In the basis (\ref{basis}), it has the explicit form \cite{Glover:2001af}
\be
\bGam^\One
= 2
\left( \begin{array}{cc}
\ia & \ib/N \\
\ic/N  & \id 
\end{array} \right)
\ee
where 
\bea
\label{defalpha}
\ia  = 
\left( {
\begin{array}{cccccc}
\tS+\tT & 0 & 0 & 0 & 0 & 0 \\
0 & \tS+\tU & 0 & 0 & 0 & 0 \\
0 & 0 & \tT+\tU & 0 & 0 & 0 \\
0 & 0 & 0 & \tT+\tU & 0 & 0\\
0 & 0 & 0 & 0 & \tS+\tU & 0\\
0 & 0 & 0 & 0 & 0 & \tS+\tT
\end{array} } \right), &&
\ib = 
\left( {
\begin{array}{ccc}
\tT-\tU & 0 & \tS-\tU \\
\tU-\tT & \tS-\tT & 0 \\
0 & \tT-\tS & \tU-\tS \\
0 & \tT-\tS & \tU-\tS \\
\tU-\tT & \tS-\tT & 0 \\
\tT-\tU & 0 & \tS-\tU \\
\end{array}
}\right)
\nonumber\\[4mm]
\ic  = 
\left( {
\begin{array}{cccccc}
\tS-\tU & \tS-\tT & 0 & 0 & \tS-\tT & \tS-\tU \\
0 & \tU-\tT & \tU-\tS & \tU-\tS & \tU-\tT& 0 \\
\tT-\tU & 0 & \tT-\tS & \tT-\tS & 0 & \tT-\tU \\
\end{array}
}\right), &&
\id = 
\left( {
\begin{array}{ccc}
2\tS & 0 & 0 \\
0 & 2\tU & 0 \\
0 & 0 & 2\tT
\end{array}
} \right)
\eea
with
\be
\tS = \log \left(-\frac{s}{Q^2}\right),\qquad\qquad 
\tT = \log \left(-\frac{t}{Q^2}\right), \qquad\qquad 
\tU = \log \left(-\frac{u}{Q^2}\right)\,.
\ee
We use \eqns{tree}{defalpha} to show
\be
\label{gammabeta}
\ic \ket{A^\Zero} = 
\left( - \frac{ 4 i K }{stu} \right) 2(sY-tX) \ones,
\quad{\rm and}\quad
\gamma \beta  \ones
= 
2 \left( X^2 + Y^2 + Z^2 \right)
\ones
\ee
where 
\be
\label{defXYZ}
X = \log \left(t \over u\right), \qquad
Y = \log \left(u \over s\right), \qquad
Z = \log \left(s \over t\right).
\ee
For consideration of the 
Regge limit $s\gg -t$, with $t<0$ held fixed, 
it is convenient to set the arbitrary factorization scale $Q^2$ equal to $-t$, 
in which case the elements of the one-loop anomalous dimension 
matrix (\ref{defalpha}) 
take the form
\bea
\label{defnewalpha}
\ia  = 
\left( {
\begin{array}{cccccc}
Z & 0 & 0 & 0 & 0 & 0 \\
0 & Z-X & 0 & 0 & 0 & 0 \\
0 & 0 & -X & 0 & 0 & 0 \\
0 & 0 & 0 & -X & 0 & 0\\
0 & 0 & 0 & 0 & Z-X & 0\\
0 & 0 & 0 & 0 & 0 & Z
\end{array} } \right), &&
\ib = 
\left( {
\begin{array}{ccc}
X  & 0 &-Y \\
-X & Z & 0 \\
0  &-Z & Y \\
0  &-Z & Y \\
-X & Z & 0 \\
X  & 0 &-Y \\
\end{array}
}\right)
\nonumber\\[4mm]
\ic  = 
\left( {
\begin{array}{cccccc}
-Y & Z & 0 & 0 & Z &-Y \\
 0 &-X & Y & Y &-X & 0 \\
 X & 0 &-Z &-Z & 0 & X \\
\end{array}
}\right), &&
\id = 
\left( {
\begin{array}{ccc}
2 Z & 0 & 0 \\
0 & - 2 X & 0 \\
0 & 0 & 0
\end{array}
} \right)
\eea
Finally,
we analytically continue the variables $X$, $Y$, and $Z$ 
to the physical region $s>0$, $u$,$t<0$, 
and then take $s \gg -t$ to obtain 
\bea
X &\to& - \log(-s/t) + \cO(t/s) 
\nonumber\\
Y &\to& i \pi + \cO(t/s)
\label{XYZlimit}\\
Z &\to& \log(-s/t) - i \pi + \cO(t/s) 
\nonumber
\eea
{}From this, we see that the leading log behavior 
of the matrices (\ref{defnewalpha}) in the limit $s\gg -t$ is 
\bea
\ia  
\to 
\left( {
\begin{array}{cccccc}
1 & 0 & 0 & 0 & 0 & 0 \\
0 & 2 & 0 & 0 & 0 & 0 \\
0 & 0 & 1 & 0 & 0 & 0 \\
0 & 0 & 0 & 1 & 0 & 0\\
0 & 0 & 0 & 0 & 2 & 0\\
0 & 0 & 0 & 0 & 0 & 1
\end{array} } \right) 
\log(-s/t),
&&
\ib
\to
\left( {
\begin{array}{ccc}
-1 & 0 & 0 \\
 1 & 1 & 0 \\
 0 &-1 & 0 \\
 0 &-1 & 0 \\
 1 & 1 & 0 \\
-1 & 0 & 0 \\
\end{array}
}\right)
\log(-s/t)
\nonumber\\[4mm]
\ic  
\to
\left( {
\begin{array}{cccccc}
 0 & 1 & 0 & 0 & 1 & 0 \\
 0 & 1 & 0 & 0 & 1 & 0 \\
-1 & 0 &-1 &-1 & 0 &-1 \\
\end{array}
}\right) 
\log(-s/t), 
&&
\id 
\to
\left( {
\begin{array}{ccc}
2 & 0 & 0 \\
0 & 2 & 0 \\
0 & 0 & 0
\end{array}
} \right)
\log(-s/t) 
\label{alphalimit}
\eea

\vfil\break

\providecommand{\href}[2]{#2}\begingroup\raggedright\endgroup

\begin{thebibliography}{10}

\bibitem{Anastasiou:2003kj}
C.~Anastasiou, Z.~Bern, L.~J. Dixon, and D.~A. Kosower, ``{Planar amplitudes in
  maximally supersymmetric Yang-Mills theory},''
  \href{http://dx.doi.org/10.1103/PhysRevLett.91.251602}{{\em Phys. Rev. Lett.}
  {\bf 91} (2003)  251602},
\href{http://arxiv.org/abs/hep-th/0309040}{{\tt arXiv:hep-th/0309040}}.
%%CITATION = HEP-TH/0309040;%%.

\bibitem{Magnea:1990zb}
L.~Magnea and G.~Sterman, ``{Analytic continuation of the Sudakov form-factor
  in QCD},''
{\em Phys. Rev.} {\bf D42} (1990)  4222--4227.
%%CITATION = PHRVA,D42,4222;%%.

\bibitem{Catani:1996jh}
S.~Catani and M.~H. Seymour, ``{The Dipole Formalism for the Calculation of QCD
  Jet Cross Sections at Next-to-Leading Order},''
  \href{http://dx.doi.org/10.1016/0370-2693(96)00425-X}{{\em Phys. Lett.} {\bf
  B378} (1996)  287--301},
\href{http://arxiv.org/abs/hep-ph/9602277}{{\tt arXiv:hep-ph/9602277}};
%%CITATION = HEP-PH/9602277;%%.
%\bibitem{Catani:1996vz}
%S.~Catani and M.~H. Seymour, 
``{A general algorithm for calculating jet cross
  sections in NLO QCD},''
  \href{http://dx.doi.org/10.1016/S0550-3213(96)00589-5}{{\em Nucl. Phys.} {\bf
  B485} (1997)  291--419},
\href{http://arxiv.org/abs/hep-ph/9605323}{{\tt arXiv:hep-ph/9605323}}.
%%CITATION = HEP-PH/9605323;%%.


\bibitem{Catani:1998bh}
S.~Catani, ``{The singular behaviour of {QCD} amplitudes at two-loop order},''
  \href{http://dx.doi.org/10.1016/S0370-2693(98)00332-3}{{\em Phys. Lett.} {\bf
  B427} (1998)  161--171},
\href{http://arxiv.org/abs/hep-ph/9802439}{{\tt arXiv:hep-ph/9802439}}.
%%CITATION = HEP-PH/9802439;%%.

\bibitem{Sterman:2002qn}
G.~Sterman and M.~E. Tejeda-Yeomans, ``{Multi-loop amplitudes and
  resummation},'' \href{http://dx.doi.org/10.1016/S0370-2693(02)03100-3}{{\em
  Phys. Lett.} {\bf B552} (2003)  48--56},
\href{http://arxiv.org/abs/hep-ph/0210130}{{\tt arXiv:hep-ph/0210130}}.
%%CITATION = HEP-PH/0210130;%%.

\bibitem{Bern:2005iz}
Z.~Bern, L.~J. Dixon, and V.~A. Smirnov, ``{Iteration of planar amplitudes in
  maximally supersymmetric Yang-Mills theory at three loops and beyond},''
  \href{http://dx.doi.org/10.1103/PhysRevD.72.085001}{{\em Phys. Rev.} {\bf
  D72} (2005)  085001},
\href{http://arxiv.org/abs/hep-th/0505205}{{\tt arXiv:hep-th/0505205}}.
%%CITATION = HEP-TH/0505205;%%.

\bibitem{Drummond:2006rz}
J.~M. Drummond, J.~Henn, V.~A. Smirnov, and E.~Sokatchev, ``{Magic identities
  for conformal four-point integrals},'' {\em JHEP} {\bf 01} (2007)  064,
\href{http://arxiv.org/abs/hep-th/0607160}{{\tt arXiv:hep-th/0607160}}.
%%CITATION = HEP-TH/0607160;%%.

\bibitem{Drummond:2007aua}
J.~M. Drummond, G.~P. Korchemsky, and E.~Sokatchev, ``{Conformal properties of
  four-gluon planar amplitudes and Wilson loops},''
  \href{http://dx.doi.org/10.1016/j.nuclphysb.2007.11.041}{{\em Nucl. Phys.}
  {\bf B795} (2008)  385--408},
\href{http://arxiv.org/abs/0707.0243}{{\tt arXiv:0707.0243 [hep-th]}}.
%%CITATION = 0707.0243;%%.

\bibitem{Drummond:2007cf}
J.~M. Drummond, J.~Henn, G.~P. Korchemsky, and E.~Sokatchev, ``{On planar gluon
  amplitudes/Wilson loops duality},''
  \href{http://dx.doi.org/10.1016/j.nuclphysb.2007.11.007}{{\em Nucl. Phys.}
  {\bf B795} (2008)  52--68},
\href{http://arxiv.org/abs/0709.2368}{{\tt arXiv:0709.2368 [hep-th]}};
%%CITATION = 0709.2368;%%.
%\bibitem{Drummond:2007au}
%J.~M. Drummond, J.~Henn, G.~P. Korchemsky, and E.~Sokatchev, 
``{Conformal Ward
  identities for Wilson loops and a test of the duality with gluon
  amplitudes},''
\href{http://arxiv.org/abs/0712.1223}{{\tt arXiv:0712.1223 [hep-th]}}.
%%CITATION = 0712.1223;%%.

\bibitem{Alday:2007hr}
L.~F. Alday and J.~M. Maldacena, ``{Gluon scattering amplitudes at strong
  coupling},'' {\em JHEP} {\bf 06} (2007)  064,
\href{http://arxiv.org/abs/0705.0303}{{\tt arXiv:0705.0303 [hep-th]}}.
%%CITATION = 0705.0303;%%.

\bibitem{Brandhuber:2007yx}
A.~Brandhuber, P.~Heslop, and G.~Travaglini, ``{MHV Amplitudes in
${\cal N}=4$ Super Yang-Mills and Wilson Loops},''
  \href{http://dx.doi.org/10.1016/j.nuclphysb.2007.11.002}{{\em Nucl. Phys.}
  {\bf B794} (2008)  231--243},
\href{http://arxiv.org/abs/0707.1153}{{\tt arXiv:0707.1153 [hep-th]}}.
%%CITATION = 0707.1153;%%.



\bibitem{Drummond:2007bm}
J.~M. Drummond, J.~Henn, G.~P. Korchemsky, and E.~Sokatchev, ``{The hexagon
  Wilson loop and the BDS ansatz for the six- gluon amplitude},''
  \href{http://dx.doi.org/10.1016/j.physletb.2008.03.032}{{\em Phys. Lett.}
  {\bf B662} (2008)  456--460},
\href{http://arxiv.org/abs/0712.4138}{{\tt arXiv:0712.4138 [hep-th]}}.
%%CITATION = 0712.4138;%%.

\bibitem{Bern:2008ap}
Z.~Bern {\em et al.}, ``{The Two-Loop Six-Gluon MHV Amplitude in Maximally
  Supersymmetric Yang-Mills Theory},''
  \href{http://dx.doi.org/10.1103/PhysRevD.78.045007}{{\em Phys. Rev.} {\bf
  D78} (2008)  045007},
\href{http://arxiv.org/abs/0803.1465}{{\tt arXiv:0803.1465 [hep-th]}}.
%%CITATION = 0803.1465;%%.

\bibitem{Drummond:2008aq}
J.~M. Drummond, J.~Henn, G.~P. Korchemsky, and E.~Sokatchev, ``{Hexagon Wilson
  loop = six-gluon MHV amplitude},''
  \href{http://dx.doi.org/10.1016/j.nuclphysb.2009.02.015}{{\em Nucl. Phys.}
  {\bf B815} (2009)  142--173},
\href{http://arxiv.org/abs/0803.1466}{{\tt arXiv:0803.1466 [hep-th]}}.
%%CITATION = 0803.1466;%%.

\bibitem{Alday:2007he}
L.~F. Alday and J.~Maldacena, ``{Comments on gluon scattering amplitudes via
  AdS/CFT},'' \href{http://dx.doi.org/10.1088/1126-6708/2007/11/068}{{\em JHEP}
  {\bf 11} (2007)  068},
\href{http://arxiv.org/abs/0710.1060}{{\tt arXiv:0710.1060 [hep-th]}}.
%%CITATION = 0710.1060;%%.

\bibitem{Anastasiou:2009kna}
C.~Anastasiou {\em et al.}, ``{Two-Loop Polygon Wilson Loops in N=4 SYM},''
  \href{http://dx.doi.org/10.1088/1126-6708/2009/05/115}{{\em JHEP} {\bf 05}
  (2009)  115},
\href{http://arxiv.org/abs/0902.2245}{{\tt arXiv:0902.2245 [hep-th]}}.
%%CITATION = 0902.2245;%%.

\bibitem{Naculich:2007ub}
S.~G. Naculich and H.~J. Schnitzer, ``{Regge behavior of gluon scattering
  amplitudes in ${\cal N} =4$ SYM theory},''
  \href{http://dx.doi.org/10.1016/j.nuclphysb.2007.10.026}{{\em Nucl. Phys.}
  {\bf B794} (2008)  189--194},
\href{http://arxiv.org/abs/0708.3069}{{\tt arXiv:0708.3069 [hep-th]}}.
%%CITATION = 0708.3069;%%.

\bibitem{DelDuca:2008pj}
V.~Del~Duca and E.~W.~N. Glover, ``{Testing high-energy factorization beyond
  the next-to- leading-logarithmic accuracy},''
  \href{http://dx.doi.org/10.1088/1126-6708/2008/05/056}{{\em JHEP} {\bf 05}
  (2008)  056},
\href{http://arxiv.org/abs/0802.4445}{{\tt arXiv:0802.4445 [hep-th]}}.
%%CITATION = 0802.4445;%%.

\bibitem{Brower:2008nm}
R.~C. Brower, H.~Nastase, H.~J. Schnitzer, and C.-I. Tan, ``{Implications of
  multi-Regge limits for the Bern-Dixon- Smirnov conjecture},''
  \href{http://dx.doi.org/10.1016/j.nuclphysb.2009.02.009}{{\em Nucl. Phys.}
  {\bf B814} (2009)  293--326},
\href{http://arxiv.org/abs/0801.3891}{{\tt arXiv:0801.3891 [hep-th]}}.
%%CITATION = 0801.3891;%%.

\bibitem{Bartels:2008ce}
J.~Bartels, L.~N. Lipatov, and A.~Sabio~Vera, ``{BFKL Pomeron, Reggeized gluons
  and Bern-Dixon-Smirnov amplitudes},''
\href{http://arxiv.org/abs/0802.2065}{{\tt arXiv:0802.2065 [hep-th]}};
%%CITATION = 0802.2065;%%.
%\bibitem{Bartels:2008sc}
%J.~Bartels, L.~N. Lipatov, and A.~Sabio~Vera, 
``{ ${\cal N}=4$ supersymmetric
  Yang Mills scattering amplitudes at high energies: the Regge cut
  contribution},''
\href{http://arxiv.org/abs/0807.0894}{{\tt arXiv:0807.0894 [hep-th]}}.
%%CITATION = 0807.0894;%%.

\bibitem{Brower:2008ia}
R.~C. Brower, H.~Nastase, H.~J. Schnitzer, and C.-I. Tan, ``{Analyticity for
  Multi-Regge Limits of the Bern-Dixon- Smirnov Amplitudes},''
\href{http://arxiv.org/abs/0809.1632}{{\tt arXiv:0809.1632 [hep-th]}}.
%%CITATION = 0809.1632;%%.

\bibitem{DelDuca:2008jg}
V.~Del~Duca, C.~Duhr, and E.~W.~N. Glover, ``{Iterated amplitudes in the
  high-energy limit},''
  \href{http://dx.doi.org/10.1088/1126-6708/2008/12/097}{{\em JHEP} {\bf 12}
  (2008)  097},
\href{http://arxiv.org/abs/0809.1822}{{\tt arXiv:0809.1822 [hep-th]}}.
%%CITATION = 0809.1822;%%.

\bibitem{Bern:1997nh}
Z.~Bern, J.~S. Rozowsky, and B.~Yan, ``{Two-loop four-gluon amplitudes in
  ${\cal N}=4$ super-Yang-Mills},''
  \href{http://dx.doi.org/10.1016/S0370-2693(97)00413-9}{{\em Phys. Lett.} {\bf
  B401} (1997)  273--282},
\href{http://arxiv.org/abs/hep-ph/9702424}{{\tt arXiv:hep-ph/9702424}}.
%%CITATION = HEP-PH/9702424;%%.

\bibitem{Smirnov:1999gc}
V.~A. Smirnov, ``{Analytical result for dimensionally regularized massless
  on-shell double box},''
  \href{http://dx.doi.org/10.1016/S0370-2693(99)00777-7}{{\em Phys. Lett.} {\bf
  B460} (1999)  397--404},
\href{http://arxiv.org/abs/hep-ph/9905323}{{\tt arXiv:hep-ph/9905323}}.
%%CITATION = HEP-PH/9905323;%%.

\bibitem{Tausk:1999vh}
J.~B. Tausk, ``{Non-planar massless two-loop Feynman diagrams with four
  on-shell legs},'' \href{http://dx.doi.org/10.1016/S0370-2693(99)01277-0}{{\em
  Phys. Lett.} {\bf B469} (1999)  225--234},
\href{http://arxiv.org/abs/hep-ph/9909506}{{\tt arXiv:hep-ph/9909506}}.
%%CITATION = HEP-PH/9909506;%%.

\bibitem{Bern:2008pv}
Z.~Bern, J.~J.~M. Carrasco, L.~J. Dixon, H.~Johansson, and R.~Roiban,
  ``{Manifest Ultraviolet Behavior for the Three-Loop Four- Point Amplitude of
  ${\cal N}=8$ Supergravity},'' {\em Phys. Rev.} {\bf D78} (2008)  105019,
\href{http://arxiv.org/abs/0808.4112}{{\tt arXiv:0808.4112 [hep-th]}}.
%%CITATION = 0808.4112;%%.

\bibitem{Naculich:2008ys}
S.~G. Naculich, H.~Nastase, and H.~J. Schnitzer, ``{Subleading-color
  contributions to gluon-gluon scattering in ${\cal N}=4$ SYM theory and
  relations to ${\cal N}=8$ supergravity},''
  \href{http://dx.doi.org/10.1088/1126-6708/2008/11/018}{{\em JHEP} {\bf 11}
  (2008)  018},
\href{http://arxiv.org/abs/0809.0376}{{\tt arXiv:0809.0376 [hep-th]}}.
%%CITATION = 0809.0376;%%.

\bibitem{MertAybat:2006wq}
S.~Mert~Aybat, L.~J. Dixon, and G.~Sterman, ``{The two-loop anomalous dimension
  matrix for soft gluon exchange},''
  \href{http://dx.doi.org/10.1103/PhysRevLett.97.072001}{{\em Phys. Rev. Lett.}
  {\bf 97} (2006)  072001},
\href{http://arxiv.org/abs/hep-ph/0606254}{{\tt arXiv:hep-ph/0606254}}.
%%CITATION = HEP-PH/0606254;%%.

\bibitem{MertAybat:2006mz}
S.~Mert~Aybat, L.~J. Dixon, and G.~Sterman, ``{The two-loop soft anomalous
  dimension matrix and resummation at next-to-next-to leading pole},''
  \href{http://dx.doi.org/10.1103/PhysRevD.74.074004}{{\em Phys. Rev.} {\bf
  D74} (2006)  074004},
\href{http://arxiv.org/abs/hep-ph/0607309}{{\tt arXiv:hep-ph/0607309}}.
%%CITATION = HEP-PH/0607309;%%.

\bibitem{Dixon:2009gx}
L.~J. Dixon, ``{Matter Dependence of the Three-Loop Soft Anomalous Dimension
  Matrix},''
\href{http://arxiv.org/abs/0901.3414}{{\tt arXiv:0901.3414 [hep-ph]}}.
%%CITATION = 0901.3414;%%.

\bibitem{Becher:2009cu}
T.~Becher and M.~Neubert, ``{Infrared singularities of scattering amplitudes in
  perturbative QCD},''
  \href{http://dx.doi.org/10.1103/PhysRevLett.102.162001}{{\em Phys. Rev.
  Lett.} {\bf 102} (2009)  162001},
\href{http://arxiv.org/abs/0901.0722}{{\tt arXiv:0901.0722 [hep-ph]}}.
%%CITATION = 0901.0722;%%.

\bibitem{Becher:2009qa}
T.~Becher and M.~Neubert, 
``{On the Structure of Infrared Singularities of
  Gauge-Theory Amplitudes},''
  \href{http://dx.doi.org/10.1088/1126-6708/2009/06/081}{{\em JHEP} {\bf 06}
  (2009)  081},
\href{http://arxiv.org/abs/0903.1126}{{\tt arXiv:0903.1126 [hep-ph]}}.
%%CITATION = 0903.1126;%%.

\bibitem{Gardi:2009qi}
E.~Gardi and L.~Magnea, ``{Factorization constraints for soft anomalous
  dimensions in QCD scattering amplitudes},''
  \href{http://dx.doi.org/10.1088/1126-6708/2009/03/079}{{\em JHEP} {\bf 03}
  (2009)  079},
\href{http://arxiv.org/abs/0901.1091}{{\tt arXiv:0901.1091 [hep-ph]}}.
%%CITATION = 0901.1091;%%.


\bibitem{Kotikov:2002ab}
A.~V. Kotikov and L.~N. Lipatov, ``{DGLAP and BFKL evolution equations in the
  ${\cal N}=4$ supersymmetric gauge theory},'' {\em Nucl. Phys.} {\bf B661}
  (2003)  19--61,
\href{http://arxiv.org/abs/hep-ph/0208220}{{\tt arXiv:hep-ph/0208220}}.
%%CITATION = HEP-PH/0208220;%%.

\bibitem{Fadin:1996tb}
V.~S. Fadin, R.~Fiore, and M.~I. Kotsky, ``{Gluon Regge trajectory in the
  two-loop approximation},''
  \href{http://dx.doi.org/10.1016/0370-2693(96)01054-4}{{\em Phys. Lett.} {\bf
  B387} (1996)  593--602},
\href{http://arxiv.org/abs/hep-ph/9605357}{{\tt arXiv:hep-ph/9605357}}.
%%CITATION = HEP-PH/9605357;%%.

\bibitem{DelDuca:2001gu}
V.~Del~Duca and E.~W.~N. Glover, ``{The high energy limit of QCD at two
  loops},'' {\em JHEP} {\bf 10} (2001)  035,
\href{http://arxiv.org/abs/hep-ph/0109028}{{\tt arXiv:hep-ph/0109028}}.
%%CITATION = HEP-PH/0109028;%%.


\bibitem{Glover:2001af}
E.~W.~N. Glover, C.~Oleari, and M.~E. Tejeda-Yeomans, ``{Two-loop QCD
  corrections to gluon gluon scattering},''
  \href{http://dx.doi.org/10.1016/S0550-3213(01)00210-3}{{\em Nucl. Phys.} {\bf
  B605} (2001)  467--485},
\href{http://arxiv.org/abs/hep-ph/0102201}{{\tt arXiv:hep-ph/0102201}}.
%%CITATION = HEP-PH/0102201;%%.

\bibitem{Armoni:2006ux}
A.~Armoni, ``{Anomalous dimensions from a spinning D5-brane},''
{\em JHEP} {\bf 11} (2006)  009,
\href{http://arxiv.org/abs/hep-th/0608026}{{\tt arXiv:hep-th/0608026}}.
%%CITATION = HEP-TH/0608026;%%.

\bibitem{Alday:2007mf}
L.~F. Alday and J.~M. Maldacena, ``{Comments on operators with large spin},''
  \href{http://dx.doi.org/10.1088/1126-6708/2007/11/019}{{\em JHEP} {\bf 11}
  (2007)  019},
\href{http://arxiv.org/abs/0708.0672}{{\tt arXiv:0708.0672 [hep-th]}}.
%%CITATION = 0708.0672;%%.

\bibitem{Dixon}
L. Dixon, private communication.

\bibitem{Bern:2006ew}
Z.~Bern, M.~Czakon, L.~J. Dixon, D.~A. Kosower, and V.~A. Smirnov, ``{The
  Four-Loop Planar Amplitude and Cusp Anomalous Dimension in Maximally
  Supersymmetric Yang-Mills Theory},''
  \href{http://dx.doi.org/10.1103/PhysRevD.75.085010}{{\em Phys. Rev.} {\bf
  D75} (2007)  085010},
\href{http://arxiv.org/abs/hep-th/0610248}{{\tt arXiv:hep-th/0610248}}.
%%CITATION = HEP-TH/0610248;%%.

\bibitem{Bern:1993kr}
Z.~Bern, L.~J. Dixon, and D.~A. Kosower, ``{Dimensionally regulated pentagon
  integrals},'' \href{http://dx.doi.org/10.1016/0550-3213(94)90398-0}{{\em
  Nucl. Phys.} {\bf B412} (1994)  751--816},
\href{http://arxiv.org/abs/hep-ph/9306240}{{\tt arXiv:hep-ph/9306240}}.
%%CITATION = HEP-PH/9306240;%%.

\bibitem{Bern:1998sc}
Z.~Bern, V.~Del~Duca, and C.~R. Schmidt, ``{The infrared behavior of one-loop
  gluon amplitudes at next-to-next-to-leading order},''
  \href{http://dx.doi.org/10.1016/S0370-2693(98)01495-6}{{\em Phys. Lett.} {\bf
  B445} (1998)  168--177},
\href{http://arxiv.org/abs/hep-ph/9810409}{{\tt arXiv:hep-ph/9810409}}.
%%CITATION = HEP-PH/9810409;%%.

\bibitem{Maitre:2005uu}
D.~Maitre, ``{HPL, a Mathematica implementation of the harmonic
  polylogarithms},'' \href{http://dx.doi.org/10.1016/j.cpc.2005.10.008}{{\em
  Comput. Phys. Commun.} {\bf 174} (2006)  222--240},
\href{http://arxiv.org/abs/hep-ph/0507152}{{\tt arXiv:hep-ph/0507152}}.
%%CITATION = HEP-PH/0507152;%%.

\bibitem{Grisaru:1982bi}
M.~T. Grisaru and H.~J. Schnitzer, ``{Bound states in ${\cal N}=8$ supergravity
  and ${\cal N}=4$ supersymmetric Yang-Mills theories},''
{\em Nucl. Phys.} {\bf B204} (1982)  267.
%%CITATION = NUPHA,B204,267;%%.

\bibitem{Green:1982sw}
M.~B. Green, J.~H. Schwarz, and L.~Brink, ``{${\cal N}=4$ Yang-Mills and ${\cal
  N}=8$ Supergravity as Limits of String Theories},''
\href{http://dx.doi.org/10.1016/0550-3213(82)90336-4}{{\em Nucl. Phys.} {\bf
  B198} (1982)  474--492}.
%%CITATION = NUPHA,B198,474;%%.

\bibitem{DelDuca:1998kx}
V.~Del~Duca and C.~R. Schmidt, ``{Virtual next-to-leading corrections to the
  impact factors in the high-energy limit},''
  \href{http://dx.doi.org/10.1103/PhysRevD.57.4069}{{\em Phys. Rev.} {\bf D57}
  (1998)  4069--4079},
\href{http://arxiv.org/abs/hep-ph/9711309}{{\tt arXiv:hep-ph/9711309}}.
%%CITATION = HEP-PH/9711309;%%.

\bibitem{Naculich:2008ew}
S.~G. Naculich, H.~Nastase, and H.~J. Schnitzer, ``{Two-loop graviton
  scattering relation and IR behavior in ${\cal N}=8$ supergravity},''
  \href{http://dx.doi.org/10.1016/j.nuclphysb.2008.07.001}{{\em Nucl. Phys.}
  {\bf B805} (2008)  40--58},
\href{http://arxiv.org/abs/0805.2347}{{\tt arXiv:0805.2347 [hep-th]}}.
%%CITATION = 0805.2347;%%.

\bibitem{Green:1987sp}
M.~B. Green, J.~H. Schwarz, and E.~Witten, ``{Superstring Theory, vol. 1},''.
  Cambridge, UK: Univ. Pr. (1987) 469 pp. (Cambridge Monographs On Mathematical
  Physics).

\end{thebibliography}
\end{document}